\documentclass[pre,pr1,reprint,twocolumn,amsmath,10pt,amssymb,aps,superscriptaddress,showkeys,showpacs]{revtex4-1}
\usepackage[utf8]{inputenc}
\usepackage{lipsum}
\usepackage{enumitem,mathtools}
\usepackage{graphicx}
\usepackage{subcaption}
\usepackage{dcolumn}
\usepackage{array}
\usepackage{bm}
\usepackage[colorlinks=true,
            linkcolor=blue,
            urlcolor  = blue,
            citecolor = blue,
            anchorcolor = blue]{hyperref}
\usepackage[
scale=0.7, marginratio={1:1, 2:3}, ignoreall,
text={7in,10in},centering,
margin=.6in,
total={6.5in,8.75in}, top=1.2in, left=0.5in, includefoot,
height=10in,a4paper,hmargin={1.5cm,0.8in},
]{geometry}
\usepackage{fancyhdr}
\usepackage{mwe}
\usepackage{relsize}
\usepackage{scalerel}
\usepackage{xparse}

\newcommand*{\Scale}[2][4]{\scalebox{#1}{$#2$}}%
\usepackage[utf8]{inputenc}
\newcommand{\be}{\begin{equation}}
\newcommand{\bea}{\begin{align}}
\newcommand{\ee}{\end{equation}}
\newcommand{\eea}{\end{align}}
\usepackage{etoolbox} 
\usepackage{xcolor}
\usepackage[capitalize]{cleveref}

\makeatletter
\appto{\appendix}{%
  \@ifstar{\def\thesection{\unskip}\def\theequation@prefix{A.}}%
          {\def\thesection{\Alph {section}}}%
}
\makeatother
\begin{document}
\title{Role of dimensions in first passage of a diffusing particle under stochastic resetting and attractive bias.}
\author{Saeed Ahmad}
\email{saeedmalik@iitb.ac.in}%
\author{Dibyendu Das}
\email{dibyendu@phy.iitb.ac.in}%
\affiliation{Physics Department, Indian Institute of Technology Bombay, Mumbai 400076, India}%
\date{\today}
\begin{abstract}
Recent studies in one dimension have revealed that the temporal advantage rendered by stochastic resetting to diffusing particles in attaining first passage, may be annulled by a sufficiently strong attractive potential. We extend the results to higher dimensions. For a diffusing particle in an attractive potential $V({R})=k {R}^n$, in general $d$ dimensions, we study the critical strength $k = k_c$ above which resetting becomes disadvantageous. The point of continuous transition may be exactly found even in cases where the problem with resetting is not solvable, provided the first two moments of the problem without resetting are known. We find the dimensionless critical strength $\kappa_{c,n}(k_c)$ exactly when 
$d/n$ and $2/n$ take positive integral values. Also for the limiting case of a box potential (representing $n \to \infty$), and the special case of a logarithmic potential $k \ln\big(\frac{R}{a}\big)$, we find the corresponding transition points $\kappa_{c,\infty}$ and $\kappa_{c,l}$ exactly for any dimension $d$. The asymptotic forms of the critical strengths at large dimensions $d$ are interesting. We show that for the power law potential, for any $n \in (0,\infty)$, the dimensionless critical strength  $\kappa_{c,n}  \sim d^{\frac{1}{n}}$ at large $d$.  For the box potential, asymptotically,  $\kappa_{c,\infty} \sim (1 - \ln(\frac{d}{2})/d)$, while for the logarithmic potential, $\kappa_{c,l} \sim d$.

\begin{description}
\item[PACS number(s)]05.40.-a,02.50.-r,02.50.Ey
\end{description}
\end{abstract}
\maketitle

\section{\label{sec1}Introduction}
In the field of stochastic processes, the general problem of arriving at a target for the first time in the course of a random evolution, is well known. Studies of such first passage in variety of theoretical models have appeared over the years~\cite{Review_redner_2001, A_Phys_Bray_Staya_2013}. It is also of great practical interest, e.g. in the context of chemical reactions~\cite{PNAS_Reuveni_Shlomi_2014_MicMenten}, proteins reaching a threshold in cells~\cite{PNAS_A_Singh_2017}, or capture of chromosomes by microtubules~\cite{PRR_indrani_PRR_2020}.  

For an undirected $1$-d random walk, a classic result is that the distribution of times of first passage is unbounded with a divergent mean \cite{feller1966guide}. But when a potential provides a bias towards the target, the distribution becomes bounded, and there is a finite mean time~\cite{Review_redner_2001} --- a recently known applied example is the lifetime distribution of microtubules within cells~\cite{Cell_Bio_Brugu_Needleman_2010}. Recently, another  important strategy called the `stochastic resetting' has been proposed and extensively studied, which too makes the mean first passage time (MFPT) finite~\cite{PRL_Evan_Staya_2011, PRL_Reuveni_2016First, Review_Evans_Satya_Reset_application_2019}. Intermittent resets to a point, eliminate very long excursions, effectively causing quicker capture at the target. Various types of resetting time distributions have been studied in the literature~\cite{PRL_APAl_Reuveni_2017FirstR,PRE_A_Nagar_S_Gupta_powerlaw_reset_2016}. The process of RNA cleavage during RNA polymerase backtracking has been theoretically modelled as a stochastic resetting~\cite{PRE_Roldan_RNA_Polymerases_2016}.

 For exponentially distributed stochastic resetting times with a rate $r$, the MFPT has a minimum value at an optimal rate $r = r_*$~\cite{PRL_Evan_Staya_2011}.  Naturally there is interest in ways of regulating this optimal resetting rate (ORR). It has been recently shown that if the two strategies of bias and resetting, as discussed above, are simultaneously imposed on a diffusing particle, a novel transition arises~\cite{PRE_Saeed_2019, J_Phys_A_Ray_Debasish_Reuveni_2019,J_Chem_Phys_Reuveni_log_Pot_2020}. An attractive potential assists in first passage suppressing the advantage of resetting.  Beyond a critical strength $k = k_c$ of the biasing potential,  the ORR $r_*$ vanishes continuously, making the stochastic resetting strategy a hindrance. A general Landau-like description has been recently developed for such transitions, which admits further possibilities of discontinuous transitions and tricritical points --- these are realised for example, in a model of biased diffusion confined within a $1-d$ box with two absorbing boundaries and an interior point of reset~\cite{PRR_Pal_Parsad_Landau_2019}. In the latter system, on varying the location  of reset, one may witness a first order transition followed by a second order transition. Similarly on varying the reset point, two successive continuous ORR vanishing transitions were found in a model of diffusion within a $1$-d box with two reflecting walls, and a partially absorbing point~\cite{J_Phys_A_Christo_Reset_Boun_2015}. The Landau-like description for the MFPT predict a mean-field exponent $\beta = 1$ near the second order point ~\cite{PRE_Saeed_2019}, and $\beta_t = 1/2$ near the tricritical point ~\cite{PRR_Pal_Parsad_Landau_2019} for the order parameter. We note that discontinuous jump of optimal parameters are also known to arise in other scenarios like resetting of discrete step random walks with  L\'evy tailed jump distributions ~\cite{PRL_Satya_Sanjib_First_2014}. In this paper we would revisit the simple scenario in which there is a single continuous transition like in ~\cite{PRE_Saeed_2019}. 
 
 
The results on diffusion and resetting at constant rate,  initially obtained for $1$-dimension~\cite{PRL_Evan_Staya_2011}, was later 
generalised to higher dimensions ~\cite{J_Phys_A_Evans_Staya_higher_d_2014}. Recently the problem of diffusion with non-instantaneous resetting has been studied in higher dimensions
\cite{Arxiv_Sokolov_Anna_Non_Inst_reset_2020}. A problem of L\'evy walk with resetting in $d$-dimensions has been studied ~\cite{PRL_Localization_transition_staya_2017}. 
A continuous transition with resetting and diffusion within a $2-d$ circle has also been studied ~\cite{PRE_Christou_reset_circle_2106}.
These works provide motivation for studying the continuous transition discussed above, where the advantage of resetting towards first passage is offset by a potential,  in general $d$-dimensions.  
With the rise of dimensions, the diffusing particle is expected to find many more directions to escape and not reach the target radius $a$. Thus, frequent resetting to a radius $R_0 > a$ would be even more necessary and effective for first passage in higher dimensions. Yet what would happen in the presence of $V(R)$ providing a drive towards the target is not clear {\it a priori}. Could it be that howsoever strong the strength $k$ of the potential is, it cannot outcompete the advantage of resetting in high enough dimensions?  The answer would certainly depend of the value of $n$ which decides the steepness of the potential.  In this paper,  we present analytically exact results for the transition point as a function of $n$ and $d$. 


 Assuming an analytic dependence of MFPT on $r$ (which is small in the vicinity of the transition), it is easy to see ORR $r_*$ vanishes exactly when \cite{PRE_Saeed_2019} the following condition is satisfied, by the first two moments of first passage time in the {\it absence} of resetting:
\be
\langle T^2 \rangle = 2 \langle T \rangle^2
\label{eq:transition}
\ee
This result may be viewed as a limit ($r_* \to 0$) of the condition $\langle T^2_{r_*} \rangle = 2 \langle T_{r_*} \rangle^2$ for ORR derived in~\cite{PRL_Reuveni_2016First},  for any problem with constant resetting rate. Note that $r_* = 0$ is both an optimal rate, as well as a transition point. The significance of Eq~(\ref{eq:transition}) is that we get information about the transition in problems in the presence of resetting, by evaluating moments in a problem in the absence of resetting. We will use this fact in this work to show that even if the full problem with resetting is intractable, the desired transition point may be obtained if we may derive the necessary quantities in the absence of resetting. 


In section II, we formulate the general problem mathematically.  In section III, we present two fully solvable cases namely the logarithmic and the uniform box potential and obtain $\kappa_{c,l}$ and $\kappa_{c,\infty}$ exactly. We also derive their asymptotic limits at large $d$. In section IV, we solve for the transition point for the general power law potential, which is the main focus of this work. Exact transition condition is derived for any $d$ and $n$, but it involves integrals which need to be numerically evaluated. This problem is shown to be further tractable for integer values of $d/n$ and $2/n$, for which we obtain full closed form results. Then the asymptotic form of $\kappa_{c,n}$ as a function of $d$ and $n$ is derived. Finally we conclude in section V. 

\section{\label{sec2} Mathematical formulation of the problem} 
We consider a diffusing point particle, with diffusivity $D$, in $d$ dimensional space, which is instantaneously absorbed if it hits a small 
target sphere of radius $a$ centred at the origin. The particle is subjected to a spherically symmetric attractive potential $V(R)=k R^n$, 
of strength $k>0$, which biases its motion towards the target.  Additionally, at a constant rate $r$, the particle position is reset back to a sphere of radius $R_0 > a$.  The first passage problem may be  developed through the standard formalism of backward Chapman-Kolmogorov equation~\cite{gard_book_2004} for the probability of survival $Q(\vec{R},t)$  up to time $t$, where $\vec{R}$ is the initial position of the particle:
\be
\frac{\partial Q}{\partial t}=
D\nabla^{2} Q- \vec{\nabla} V.\vec{\nabla} Q-rQ+rQ_0.
\label{eq:general_CK}
\ee
Here  $Q_0 \equiv Q(\vec{R_0},t)$. The initial condition $Q(|\vec{R}|,0)=1$, boundary condition $Q(|\vec{a}|,t)=0$, the potential $V(R)$,  and the reset position $|\vec{R}_0|$ are all isotropic in space. Hence the function $Q$, the gradient, and Laplacian operators in Eq.(\ref{eq:general_CK}) depend on space only through the radial distance $R$. The equation for its Laplace transform $q(R,s)=\int^{\infty}_{0}\mathrm{d}t Q(R,t)e^{-st}$ is: 
\be
\frac{d^2  q}{d R^2}+\bigg(\frac{d-1}{R}- \frac{V^{\prime}}{D}\bigg)\frac{d q}{d R}-\alpha^2 q=\frac{-1-rq_0}{D},
\label{eq:general_CK_in_LT}
\ee
where $q_0 \equiv q(R_0,s)$, $V^{\prime} = dV/dR$, and $\alpha=\sqrt{\frac{r+s}{D}}$. An exact solution 
of $q(R,s)$ is desirable as it leads to the mean first passage time (MFPT, with resetting) $\langle T_r \rangle = q(R_0,s) |_{s=0}$ for $R = R_0$.
Further, obtaining the optimal rate $r = r_*(k)$ at which $\langle T_r \rangle$ is a minimum, one is led to the critical potential strength $k_c$ and its dimensionless counterpart $\kappa_{c,n}$ at which the benefit of resetting vanishes, i.e. $r_* \to 0$. For analytically difficult cases, the MFPT, the ORR and the transition strength for any $n$ and any small $d$ may also be obtained using the numerical technique developed in Ref-\cite{PRE_Saeed_2019}.


 As we will see below, for arbitrary dimensions $d$, the solution of $q(R,s)$ in closed form is possible for the special cases of (i) the logarithmic potential $V(R) = k \ln (R/a)$, and (ii) the box potential: $V(R \in [a, R_m)) = 0$ and $V(R \geq R_m)=\infty$. This however is challenging for the potential $V(R) = k R^n$ in general $d$. Attempting transformations like $q(R,s)=\theta^j w(R,s)$ and $R=c \theta^i$, with choices of constants $i,j,c$, we could not reduce Eq.(\ref{eq:general_CK_in_LT}) to any standard differential equation \cite{Arfken_7ed_book}.  
Of course for $d=1$, the cases of $n=1$ \cite{PRE_Saeed_2019,J_Phys_A_Ray_Debasish_Reuveni_2019} and $n=2$ \cite{PRE_Saeed_2019} were solved earlier. Yet the transition point $k_c$ 
may be solved even in such cases lacking solvability of $q(R,s)$, following Eq.~(\ref{eq:transition}). The two moments $\langle T \rangle=-\int^{\infty}_0t\frac{\partial Q}{\partial t}dt=\int^{\infty}_0 Qdt$ and $\langle T^2 \rangle=-\int^{\infty}_0t^2\frac{\partial Q}{\partial t}dt=2\int^{\infty}_0 tQdt$ in the absence of resetting, may be solved from the equations: 
\begin{align}
&D\nabla^{2} \langle T \rangle- \vec{\nabla} V.\vec{\nabla} \langle T \rangle=-1, 
\label{eq:1st_Moment_without_r} \\
&D\nabla^{2} \langle T^2 \rangle- \vec{\nabla} V.\vec{\nabla} \langle T^2 \rangle=-2\langle T \rangle,
\label{eq:2nd_Moment_without_r}
\end{align}
The above equations are obtained from Eq.~(\ref{eq:general_CK}) by setting $r=0$, multiplying by $t$ and $t^2$ respectively, and then integrating over $t$. 
 
\section{\label{sec3} Two fully solvable cases in arbitrary dimensions $d$}
Interestingly for the two potentials mentioned above, namely (i) the  logarithmic and (ii) the box, the  Eq.~(\ref{eq:general_CK_in_LT}) reduces to the same form as below: 
\be
\frac{\partial^2  q}{\partial R^2}+\bigg(\frac{1-2b}{R}\bigg)\frac{\partial q}{\partial R}-\alpha^2 q=\frac{-1-rq_0}{D},
\label{eq:d_dim_LT_n0}
\ee
The constant $b$ takes different values in the two cases which would be denoted below by $\nu(d,k)$ and $\mu(d)$, respectively.   The Eq.(\ref{eq:d_dim_LT_n0}) becomes the modified Bessel equation~\cite{Abramowitz_handbook}, as the constant on the right hand side is absorbed in a redefined $q$. 
The equations of the moments without resetting (Eqs.(\ref{eq:1st_Moment_without_r}),(\ref{eq:2nd_Moment_without_r})), for both these potentials, may be rewritten in common forms as: 
\begin{align}
&\frac{d^2  \langle T \rangle}{d R^2}+\bigg(\frac{1-2b}{R}\bigg)\frac{d \langle T \rangle}{d R}=-\frac{1}{D},
\label{eq:d_mean_time_n=0} \\
&\frac{d^2  \langle T^2 \rangle}{d R^2}+\bigg(\frac{1-2b}{R}\bigg)\frac{d \langle T^2 \rangle}{d R}=-\frac{2\langle T \rangle}{D},
\label{eq:d_2nd_mom_time_n=0}
\end{align}

\subsubsection{\label{Sub:section_n=0}\textbf{The Logarithmic Potential}}
For the potential $V(R)=k\ln(R/a)$, $q(R,s)$ satisfies Eq.(\ref{eq:d_dim_LT_n0}) along with the boundary condition $q(a,s) = 0$,
where $b\equiv \nu (d,k) =1-\big(\frac{d}{2}-\frac{k}{2D}\big)$. Note that the results thus acquire an interesting symmetry for this potential. Two separate problems with dimensions and potential strengths say $(d_1,k_1)$ and $(d_2,k_2)$ respectively, would nevertheless have identical values of $\langle T_r \rangle$ and hence the transition point, provided they have the same values of $\big(\frac{d_1}{2}-\frac{k_1}{2D}\big) = \big(\frac{d_2}{2}-\frac{k_2}{2D}\big) = 1 - \nu$.  All answers thus depend on $\nu$ and not separately on $d$ and $k$. The solution is 
\be
q(R,s)=\frac{a^{\nu}K_{\nu}(\alpha a)-R^{\nu}K_{\nu}(\alpha R)}{rR_0^{\nu}K_{\nu}(\alpha R_0)+sa^{\nu}K_{\nu}(\alpha a)}.
\label{eq:Survival_Prob_n=0}
\ee
By setting initial radius same as the resetting radius, $i.e$ $R=R_0$, the MFPT in terms of dimensionless quantities $\nu$, $z=\alpha_0 R_0 = R_0 \sqrt{r/D}$, and $\epsilon = \frac{a}{R_0}$ is:
\be
\langle T_r \rangle = q(R_0,s) |_{s=0} = \frac{R^2_0}{Dz^2}\bigg[\epsilon^{\nu}\frac{K_{\nu}(\epsilon z)}{K_{\nu}(z)}-1\bigg].
\label{eq:MFPT_n=0}
\ee
\begin{figure}[ht!]
\centering
 \includegraphics[width = .450\textwidth,height=0.2\textheight]{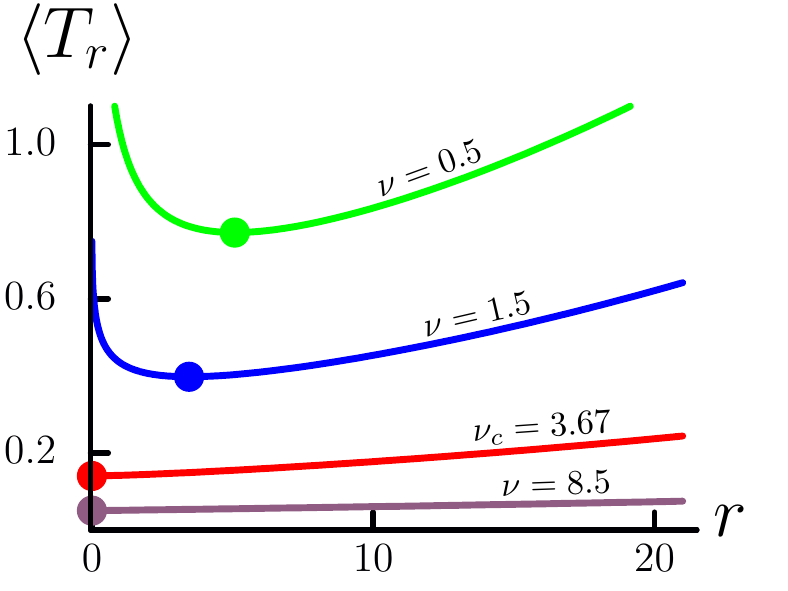}
      \caption{We show $\langle T_r \rangle $ against $r$ for the logarithmic potential. The curves are for $\epsilon = 0.5$ and $D=0.5$. The minimum values of $\langle T_r \rangle$ (and corresponding $r_*$) are shown by filled dots. It is seen that $r_* = 0$ for $\nu \geq \nu_c$.
      }
      \label{fig:MFPT_T_r_Log}
\end{figure}
In Fig.~(\ref{fig:MFPT_T_r_Log}), we have plotted $\langle T_r \rangle$ against $r$ for various of $\nu$ --- the optimal rate $r_*$ corresponding to 
the filled dots vanishes for $\nu \geq \nu_c$.  The critical value $\nu_c$ only depends on $\epsilon$ as will be shown below. The optimal rate $r_*$ or equivalently $z_*=\sqrt{\frac{r_*}{D}} R_0$ may be obtained by setting $\frac{\partial \langle T_r \rangle }{\partial r}|_{r=r_*}=0$, leading to the equation: 
\be
\frac{z_*\epsilon^{\nu+1}K_{\nu-1}(\epsilon z_*)-2K_{\nu}(z_*)}{z_*K_{\nu-1}(z_*)-2K_{\nu}(z_*)}=\epsilon^{\nu}\frac{K_{\nu}(\epsilon z_*)}{K_{\nu}(z_*)}.
\label{eq:ORR_n=0}
\ee
By setting $z_* \to 0$, we may obtain $\nu_c$ (see Appendix~(\ref{Sub:secn=0})). Yet another elegant approach is to look at the problem 
without resetting and solve Eqs.(\ref{eq:d_mean_time_n=0}) and (\ref{eq:d_2nd_mom_time_n=0}), leading to (see Appendix~(\ref{Sub1:secn=0}) for details):
\begin{align}
\langle T \rangle&=\frac{R^2(1-\epsilon^2)}{4D(\nu-1)},
\label{eq:d2_mean_time_n=0} \\
\langle T^2 \rangle&=\frac{R^4(1-\epsilon^2)}{16D^2(\nu-1)^2}\bigg(\frac{(\nu-1)(1+\epsilon^2)-2\epsilon^2(\nu-2)}{\nu-2}\bigg). 
\label{eq:d3_2nd_moment_time_n=0}
\end{align}
Then using the equality of moments at the transition (Eq.(\ref{eq:transition})), we have $\nu_c = (3-\epsilon^2)/(1-\epsilon^2)$.  For Fig.~(\ref{fig:MFPT_T_r_Log}), with $\epsilon = 0.5$, this gives $\nu_c = 3.67$.  From the expression of $\nu_c$, we have the dimensionless scaled potential strength $\kappa_{c, l}=\frac{k_c}{D}$ given by,
\be
\kappa_{c, l}=\frac{4}{1-\epsilon^2}+d
\label{eq:transition_Log1}
\ee
Thus $\kappa_{c,l}$ linearly rises with dimension $d$. This implies that as $d$ grows, the potential strength has to become unbounded, to offset the advantage of resetting.


\subsubsection{\label{Sub:section_n=Infy}\textbf{Box Potential}} 
The box potential has $V(R)=0$ for $R< R_m$ and $V(R)\to\infty$ for $R\geq R_m$. This may be thought as the limiting case of the general potential $V(R)=kR^n$ --- replacing $k=k_0/R_m^n$, we have $V(R)=k_0\big(\frac{R}{R_m}\big)^n$ which becomes the box potential in the limit $n \to \infty$. Like the logarithmic potential, $q(R,s)$ satisfies Eq.(\ref{eq:d_dim_LT_n0}), but now with a different $b\equiv\mu(d) = 1-\frac{d}{2}$. The two boundary conditions are: $q(R=a,s)=0$ (absorbing) and  $\frac{\partial q}{\partial R}|_{R=R_m}=0$ (reflecting). The potential gets stronger with diminishing $R_m$, as smaller confining space would affect quicker first passage. Accordingly, the dimensionless potential strength is $\kappa_{\infty} =  \frac{R_0}{R_m}$. Note that unlike the logarithmic potential, the constant $b \equiv \mu$ is only dependent on the dimension $d$ and not the potential strength (through $R_m$); hence there is no special symmetry in the results. We have 
\begin{align}
&q(R,s)=\bigg(\frac{rq_{0}+1}{r+s}\bigg)\bigg[1- \nonumber \\
& \bigg(\frac{R}{a}\bigg)^{\mu}
\bigg(\frac{I_{\mu}(\alpha R)K_{\mu-1}(\alpha R_m)+K_{\mu}(\alpha R)I_{\mu-1}(\alpha R_m)}{I_{\mu}(\alpha a)K_{\mu-1}(\alpha R_m)+K_{\mu}(\alpha a)I_{\mu-1}(\alpha R_m)}\bigg)\bigg].
\label{eq:Survival_n=infty}
\end{align}
By setting initial point same as the resetting point ($R=R_0$), the above gives MFPT: 
\begin{align}
&\langle T_r \rangle =q(R_0,s) |_{s=0}  = \nonumber \\ 
& \frac{1}{r}\bigg[\bigg(\frac{a}{R_0}\bigg)^{\mu} 
\bigg(\frac{I_{\mu}(\epsilon z)K_{\mu-1}(\frac{z}{\kappa_{\infty}})+K_{\mu}(\epsilon z)I_{\mu-1}(\frac{z}{\kappa_{\infty}})}{I_{\mu}(z)K_{\mu-1}(\frac{z}{\kappa_{\infty}})+K_{\mu}(z)I_{\mu-1}(\frac{z}{\kappa_{\infty}})}\bigg)-1\bigg].
\label{eq:MFPT_n=infty}
\end{align} 
The equation $\frac{\partial \langle T_r \rangle }{\partial r}|_{r=r_*}=0$ satisfied by the optimal resetting rate $r_*$ is a bit lengthy and is shown in Appendix~(\ref{Sub:secn=infty}). In Fig.~(\ref{fig:MFPT_T_r_Box}), we plot $\langle T_r \rangle$ versus $r$ using Eq~(\ref{eq:MFPT_n=infty}) for a few dimensions at the corresponding critical potential strengths $\kappa_{\infty} = \kappa_{c,\infty} = \frac{R_0}{R_{m,c}}$ such that $r_* = 0$.  We see that $\kappa_{c,\infty}$ stays finite with rising dimensions.  

\begin{figure}[hb!]
      \includegraphics[width = 0.450\textwidth,height=0.20\textheight]{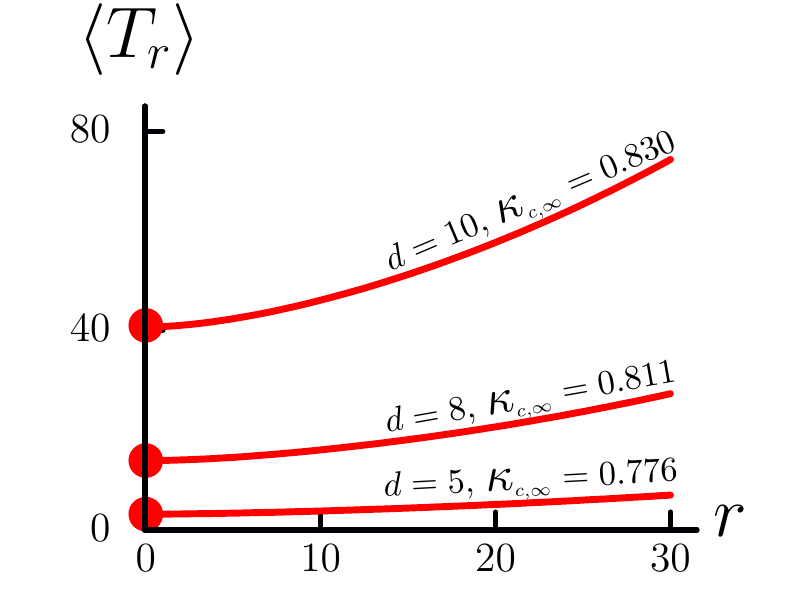}
    \caption{We show $\langle T_r \rangle$ at critical potential strengths $\kappa_{c,\infty}$ corresponding to a few dimensions $d = 5, 8$ and $10$. The red dots are minima of the curves at $r_* = 0$. Here $a=0.5$, $R_0=1.0$ and $D=0.5$.
    }
    \label{fig:MFPT_T_r_Box}
\end{figure}

\begin{figure}[ht!]
\centering
\includegraphics[width = .50\textwidth,height=0.25\textheight]{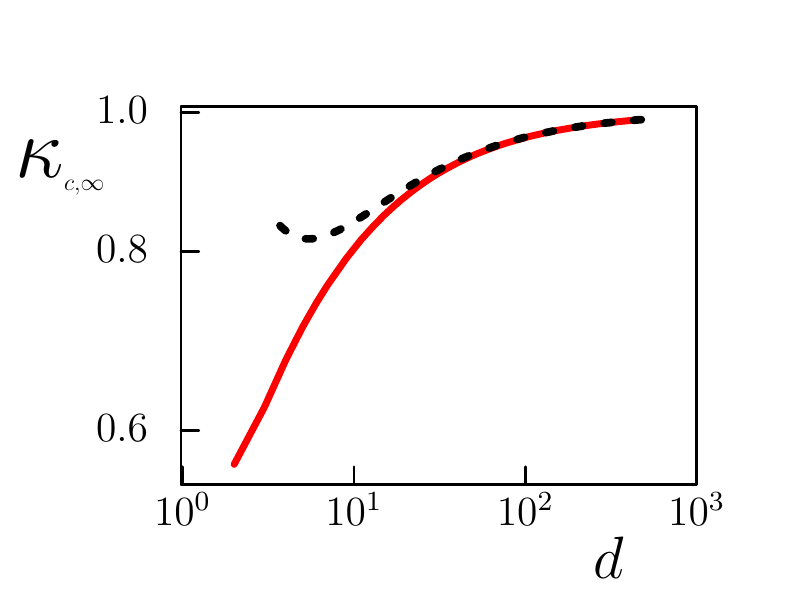}
\caption{We show $\kappa_{c, \infty}$ (solid line) from Eq~(\ref{eq:transition_box}) and $\kappa^{\rm{aym}}_{c, \infty}$ (dashed line) from  Eq~(\ref{eq:Box_Pot_Large_d}), matching each other at large $d$. 
}
\label{fig:Box_Pot_Large_d}
\end{figure}

The two moments of first passage time for the problem without resetting are respectively the following (see details of calculation in Appendix~(\ref{Sub2:secn=infty})) for $d \neq 2$:
\be
\langle T \rangle=\bigg(\frac{R^2 \kappa^{2(\mu-1)}_{\infty}\epsilon^{2\mu}}{4D\mu(\mu-1)}\bigg)\bigg[1-\epsilon^{-2\mu}\big(1+\mu(\epsilon^2-1) \kappa^{2(1-\mu)}_{\infty}\big)\bigg],
\label{eq:d2_mean_time_n=Infy}
\ee 
\be
\begin{split}
  \langle T^2 \rangle=\bigg(\frac{R^4\kappa^{4(\mu-1)}_{\infty}\epsilon^{4\mu}}{8D^2\mu^2(\mu-1)^2}\bigg)\bigg[1+\epsilon^{-2\mu}\big(\mu \kappa^{2(1-\mu)}_{\infty} - 1+\\
    \frac{2(\mu-1)^2\kappa^{-2\mu}_{\infty}}{\mu-2}\big)+O(\epsilon^{2(1-\mu)})\bigg],
\label{eq:d3_mean_time_n=Infy}
\end{split}
\ee
Following Eq.~(\ref{eq:transition}) we obtain (see Appendix~(\ref{Sub2:secn=infty})) the critical strength $\kappa_{c,\infty}$ satisfying:
\be
\frac{2(1-\mu)^2\kappa^{-2\mu}_{c, \infty}}{2-\mu}=1-\mu \kappa^{2(1-\mu)}_{c, \infty}
\label{eq:transition_box}
\ee
The above equation is a polynomial in $\kappa_{c,\infty}$ of order $2(1-\mu) \equiv d$. 
The physically acceptable root has to be $< 1$, as $R_0 < R_m$. We plot the numerical solution 
of Eq~(\ref{eq:transition_box}) in Fig~(\ref{fig:Box_Pot_Large_d}) and we see that it approaches $1$ 
at large $d$. This asymptotic approach may be analytically obtained by expanding in large $d$ and small
value of $(1- \kappa_{c,\infty})$ as shown in  Appendix~(\ref{Sub2:secn=infty}), leading to the following expression:
\be
\kappa^{\rm{asym}}_{c, \infty} \sim 1-\frac{1}{d} \ln{\big(\frac{d}{2}\big)}
\label{eq:Box_Pot_Large_d}
\ee 
The asymptotic form  matches well with the exact form as $d \to \infty$, as shown in Fig~(\ref{fig:Box_Pot_Large_d}). In contrast to the logarithmic potential, finite $\kappa_{c,\infty}$ for the box potential implies that the strategy of resetting may be made irrelevant quite easily by reducing the box size (enhancing potential strength), howsoever large the dimension $d$ may be. We would like to see now what happens in general for the potential forms with finite values of $n$. 

\section{\label{Sub:secn=infty}Transition point for the power law potential}
As noted earlier, for $V(R)=kR^n$ with $k>0$ and $n>0$, finding explicit solution for $q(R,s)$ from Eq~(\ref{eq:general_CK_in_LT}) is difficult. Yet if we are just interested in the critical strength at the transition point, namely $\kappa_{c,n} = \big( \frac{k_c}{D}\big)^{\frac{1}{n}} R_0$,  progress may be made using the relationship of Eq.~(\ref{eq:transition}).  Note that $R_0=R$ for our subsequent discussion, as the initial and resetting radii are same. The two equations 
Eq~(\ref{eq:2nd_Moment_without_r}) now take the forms: 
\begin{align}
&\frac{d^2  \langle T \rangle}{d R^2}+\bigg(\frac{d-1}{R}-n\bigg(\frac{k}{D}\bigg)R^{n-1}\bigg)\frac{d \langle T \rangle}{d R}=-\frac{1}{D},
\label{eq:d_mean_time_n} \\
&\frac{d^2  \langle T^2 \rangle}{d R^2}+\bigg(\frac{d-1}{R}-n\bigg(\frac{k}{D}\bigg)R^{n-1}\bigg)\frac{d \langle T^2 \rangle}{d R}=-\frac{2\langle T \rangle}{D}
\label{eq:d_seconed_mean_time_n}
\end{align}
The solution for $\xi=\frac{d \langle T \rangle}{d R}$ satisfying a first order differential equation (from Eq~(\ref{eq:d_mean_time_n})) is  (see Appendix~(\ref{App3Sub1:secn=n_transition})): 
\be
\xi(R)=\frac{R}{nD}\frac{e^{\frac{kR^n}{D}}}{\big(\frac{kR^n}{D}\big)^{d'}}\Gamma(d',\frac{kR^n}{D}),
\label{eq:xi_1}
\ee
where, we defined $d'=\frac{d}{n}$.

The absorbing condition $\langle T \rangle|_{R=a}=0$ implies that the mean time $\langle T \rangle$ can be obtained by integrating $\xi(R)$ from $a$ to $R$. Using the dimensionless variable $x=\frac{k{R^{'}}^n}{D}$ and noting that $\kappa_{n}^{n} = \frac{k{R}^n}{D}$, and $\frac{k{a}^n}{D} = \epsilon^n \kappa_n^n$,  we have (Appendix~(\ref{App3Sub1:secn=n_transition})): 
\be
\begin{split}
  \langle T \rangle=\frac{1}{n^2D}\bigg(\frac{D}{k}\bigg)^{\frac{2}{n}}\int_{\epsilon^n \kappa_n^n}^{\kappa_n^n}x^{\frac{2}{n}-d'-1}\Gamma(d',x)e^{x}dx \textcolor{white}{-}\\
  \equiv\frac{1}{n^2D}\bigg(\frac{D}{k}\bigg)^{\frac{2}{n}}\psi_{1}(d',n,\kappa_n^n)-\phi_{1}(d',n,\epsilon^n \kappa_n^n)
\label{eq2:d_mean_time_n}
\end{split}
\ee
where $\psi_{1}(d',n,x)$ is the indefinite integral $\int x^{\frac{2}{n}-d'-1} e^{x} \Gamma(d', x)dx$.  The quantity $\psi_1(d',n,\kappa_n^n)$ is a function of the variable $R$ (through $\kappa_n^n$). Here $\phi_{1}$ is a constant independent of $R$.

 Noting that the Eqs~(\ref{eq:d_seconed_mean_time_n}) and (\ref{eq:d_mean_time_n}) are of the same form except for the right hand side, we have: 
\be
\frac{d \langle T^2 \rangle}{d R}= \frac{2}{D}R^{1-d}e^{\frac{k}{D}R^n}\int^{\infty}_{R} \langle T \rangle R'^{d-1} e^{-\frac{kR'^n}{D}} dR',
\label{eq1:d_second_mean_time_n}
\ee
which on substituting $\langle T \rangle$ (from Eq.~(\ref{eq2:d_mean_time_n})) yields, 
\be
\begin{split}
 \frac{d \langle T^2 \rangle}{d R}= \frac{2}{n^3D^2}\bigg(\frac{D}{k}\bigg)^{d'+\frac{2}{n}} R^{1-d}e^{\frac{kR^n}{D}}\psi_{2}(d',n,\kappa_n^n)\\
\textcolor{white}{------}-2\phi_{1}(d',n,\epsilon^n \kappa_n^n)\xi(R),
\label{eq2:d_seconed_mean_time_n}
\end{split}
\ee
where $\psi_{2}(d',n,x)=\int^{\infty}_{x}x^{'(d'-1)}e^{-x'}\psi_{1}(d',n,x')dx'$.
Integrating over $R$, we get 
\be
\begin{split}
 \langle T^2 \rangle =  \frac{2}{n^4D^2}\bigg(\frac{D}{k}\bigg)^{\frac{4}{n}}\int_{\epsilon^n \kappa_n^n}^{\kappa_n^n}x^{\frac{2}{n}-d'-1}\psi_{2}(d',n,x)e^{x}dx\\
 \textcolor{white}{---}-2\phi_{1}(d',n,\epsilon^n \kappa_n^n)\langle T \rangle. 
\label{eq3:d_seconed_mean_time_n}
\end{split}
\ee
Using Eqs~(\ref{eq2:d_mean_time_n}) and (\ref{eq3:d_seconed_mean_time_n}) in Eq.~(\ref{eq:transition}) for the transition point, we finally obtain the equation 
satisfied by the dimensionless critical potential strength $\kappa_{c,n}$: 
\be
\begin{split}
 \int_{\epsilon^n {\scaleto{\kappa^{n}_{\scaleto{c, n}{2pt}}}{10pt}}}^{{\scaleto{\kappa^{n}_{\scaleto{c, n}{2pt}}}{10pt}}}x^{\frac{2}{n}-d'-1}\psi_{2}(d',n,x)e^{x}dx\textcolor{white}{-------}\\
 =\psi_{1}(d',n,{\scaleto{\kappa^{n}_{\scaleto{c, n}{2pt}}}{10pt}}) \int_{\epsilon^n {\scaleto{\kappa^{n}_{\scaleto{c, n}{2pt}}}{10pt}}}^{{\scaleto{\kappa^{n}_{\scaleto{c, n}{2pt}}}{10pt}}}x^{\frac{2}{n}-d'-1}\Gamma(d',x)e^{x}dx
\end{split}
\label{eq3:d_tansition_n}
\ee

 The above Eq.~(\ref{eq3:d_tansition_n}) is valid for any $d$ and $n$, and thus maybe numerically solved to 
obtain $\kappa_{c,n}$. But as it involves multiple functions which are themselves integrals, its numerical solution 
in high dimensions  are not likely to be very precise.  So further analytical simplification is desirable, if possible.  
We note that the function $\Gamma(d',x)$ has a series representation with finite number of terms if $d' = d/n = l$ (a positive integer): 
\be
\Gamma(l,x) = \Gamma(l) e^{-x} \sum_{j=0}^{l-1} \frac{x^j}{j!}
\label{gamma_int}
\ee  

\begin{figure}[ht!]
\centering
\includegraphics[width = .52\textwidth,height=0.25\textheight]{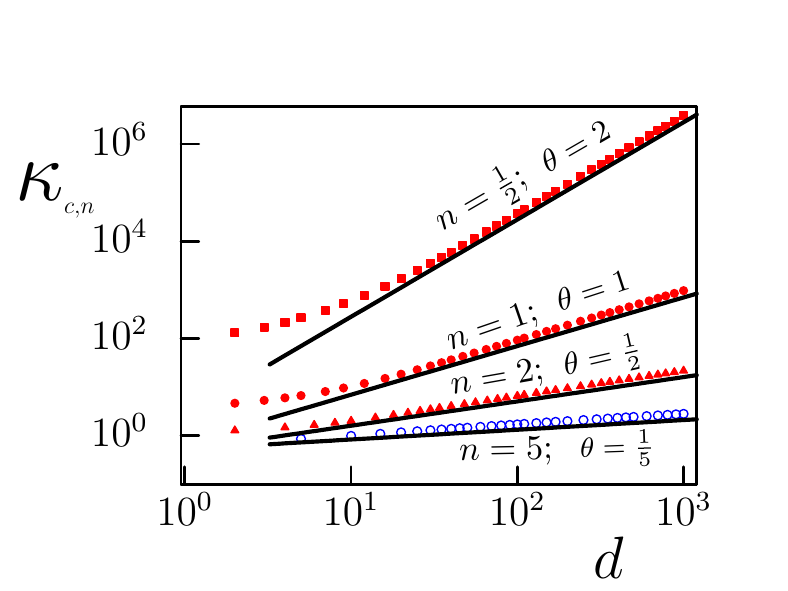}
\caption{We show ${\scaleto{\kappa_{\scaleto{c, n}{3pt}}}{8pt}}$ vs. $d$ in the log-log scale for $n=\frac{1}{2}, 1, 2$ (upper curves in red symbols) and $5$ (curve in blue symbols). The parameters used were $a=0.5$, $R_0=1.0$ and $D=0.5$. We have put thick black lines with ${\scaleto{\kappa_{\scaleto{c, n}{3pt}}}{8pt}}\sim d^{\theta}$ against each data, with $\theta = 1/n$.}
\label{fig:n_1_4_Pot_Large_d}
\end{figure}

Using the above Eq.~(\ref{gamma_int}) for the special cases of $d'$ being a positive integer, we obtain exact closed form solutions 
for $\langle T \rangle$ (with the associated $\psi_1$ function), and $d\langle T^2 \rangle/dR$ (with the associated $\psi_2$ function) -- see Eqs.~(\ref{App_eq2:d_mean_time_integer_dp}) and (\ref{App_eq3:d_seconed_mean_time_integer_dp}) in Appendix~(\ref{App3Sub2:secn=n_int_trans}) for the explicit forms.  But the expression of $d\langle T^2 \rangle/dR$ contains a 
further complication (Eq~\ref{App_eq3:d_seconed_mean_time_integer_dp}). It involves functions  $\Gamma(i + \frac{2}{n},x)$ with new arguments, where 
$i$ is non-negative integer and $2/n$ is not necessarily an integer. Thus Eq.~(\ref{gamma_int}) cannot be used for these
functions to obtain $\langle T^2 \rangle$ in closed form. Thus beyond this point,  Eq.~(\ref{eq2:d_seconed_mean_time_n}) has to be numerically 
integrated to obtain   $\langle T^2 \rangle$ and the using Eq.~(\ref{eq:transition}) one may further obtain $\kappa_{c,n}$.  We demonstrate  this procedure to 
obtain $\kappa_{c,5}$ for $n = 5$.  For all dimensions $d$ which are integer multiples of $5$, we show the critical potential strengths in Fig.~\ref{fig:n_1_4_Pot_Large_d} (see lowest lying data in blue symbols). 
It seems to follow a power law in the log-log plot.  Further progress can be made by making yet another special choice.

 If we now choose $2/n$ to be an integer (for example $n=1$, $2$ or $1/2, 1/3, 2/3$ etc.), in addition to $d'$ being an integer, 
then  Eq.~(\ref{gamma_int}) may be used for the functions $\Gamma(i + \frac{2}{n},x)$ and we obtain full exact solution 
for $\langle T^2 \rangle$ (see Eq.~(\ref{App_eq5:d_seconed_mean_time_integer_dp})).  Now we are not restricted by any precision problem of numerical integration, and Eq.~(\ref{eq3:d_tansition_n}) exactly gives $\kappa_{c,n}$.  
For the three cases $n=1$, $2$ and $1/2$,
we show plot of $\kappa_{c,n}$ obtained in this way, for different $d$ (with integer $d'$) in Fig.~\ref{fig:n_1_4_Pot_Large_d} (data is red symbols). We clearly see that the curves approach power laws for large $d$.  
Although these observations are made for special choices of $n$ and $d$, we would now present analytical arguments in the 
next section and show that in general, indeed $\kappa_{c,n} \sim d^{\theta}$ at large $d$, with $\theta = 1/n$.

\subsubsection{\label{Sub:section_n}\textbf{The transition point at large dimensions $d$}}
For the discussion in this section, we would refer to the left hand side of Eq.~(\ref{eq3:d_tansition_n}) as LHS and its right and side as 
RHS. We would show that at large $d$, for $\kappa_n^n \ll d'$ we have LHS $>$ RHS, while for $\kappa_n^n \gg d'$ we have LHS $<$ RHS. 
A crucial fact is that $\Gamma(d',x)$ switches its behaviour across $x = d'$:

\be
\Gamma(d',x)\sim
\begin{cases}
    \Gamma(d'), & \text{if $x \ll d'$}.\\
    x^{d'-1}e^{-x}, & \text{if $x \gg d'$}.
\end{cases}
\label{Gamma}
\ee

(i) For $\kappa^{n}_n \ll d'$, using the relevant limit of $\Gamma(d',x)$ from Eq.~(\ref{Gamma}), we obtain the following approximate values of 
 the functions $\psi_{1}(d',n,x)$ and  $\psi_{2}(d',n,x)$ defined in the previous section --- see Appendix~(\ref{App3Sub3:secn=n_transition}) for detailed derivation: 
\be
\begin{split}
\psi_{1}(d',n,x)\sim - \Gamma(d')\frac{x^{\frac{2}{n}-d'}}{d' - \frac{2}{n}}, \\
{\rm and} ~ \psi_{2}(d',n,x)\sim -\frac{\Gamma(d')\Gamma\big(\frac{2}{n}\big)}{d'-\frac{2}{n}}
\end{split}
\ee  
As a result, 
\be
{\rm RHS}\sim -\bigg(\frac{\Gamma(d'){\scaleto{\kappa^{\scaleto{2-d}{4pt}}_{\scaleto{n}{2pt}}}{12pt}}}{d'-\frac{2}{n}}\bigg)^{2}(\epsilon^{2-d}-1)
\label{eq:RHS_small_k}
\ee
and 
\be
{\rm LHS}\sim -\frac{\Gamma(d') \Gamma\big(\frac{2}{n}\big)}{\big(d'-\frac{2}{n}\big)^2}{\scaleto{\kappa^{\scaleto{2-d}{4pt}}_{\scaleto{n}{2pt}}}{12pt}}(\epsilon^{2-d}-1),
\label{eq:LHS_small_k}
\ee
both of which are negative. We see that RHS has an extra factor of $\Gamma(d') \kappa_n^{2-d} \approx \kappa_n^2 ~ \exp[d' \ln \big(\frac{d'}{e\kappa_n^n}\big)] \gg 1$ at large $d$ and $\kappa_n^n \ll d'$, and hence RHS is more negative than LHS. Thus we get LHS $ > $ RHS.

(ii) For $\kappa_n^n \gg d'$, using the form of $\Gamma(d',x)$ for $x \gg d'$, we get (Appendix~(\ref{App3Sub3:secn=n_transition}))
\be
\psi_{1}(d',n,x)\sim 
\begin{cases}
   \ln(x) , & \text{if $n=2$}.\\
     \frac{x^{\frac{2}{n}-1}}{\frac{2}{n}-1}, & {\rm otherwise}. 
\end{cases}
\ee
Although $\epsilon < 1$, we need to consider $\epsilon^n \kappa_n^n \gg d'$ just like $\kappa_n^n \gg d'$,
and thus working with large $x$ limit for all the cases, we get 
\be 
\psi_{2}(d',n,x)\sim
\begin{cases}
   x^{d'-1}e^{-x}\ln(x) , & \text{if $n=2$}.\\
     \frac{x^{d'+\frac{2}{n}-2}e^{-x}}{\frac{2}{n}-1}, & {\rm otherwise}, 
\end{cases}
\ee
\be
{\rm RHS} \sim
\begin{cases}
   4\ln({\scaleto{\kappa_{\scaleto{2}{2pt}}}{8pt}})\ln(1/\epsilon) , & \text{if $n=2$}.\\
     \bigg(\frac{{\scaleto{\kappa^{\scaleto{2-n}{4pt}}_{\scaleto{n}{2pt}}}{12pt}}}{\frac{2}{n}-1}\bigg)^{2}(1-\epsilon^{2-n}), & {\rm otherwise}, 
\end{cases}
\label{eq:RHS_large_k}
\ee
and 
\be
{\rm LHS}\sim
\begin{cases}
  4\ln({\scaleto{\kappa_{\scaleto{2}{2pt}}}{8pt}})\ln(1/\epsilon)\bigg(1+\frac{\ln(\epsilon)}{2\ln({\scaleto{\kappa_{\scaleto{2}{2pt}}}{6pt}})}\bigg), & \text{if $n=2$}.\\
  \frac{1}{2}\bigg(\frac{{\scaleto{\kappa^{\scaleto{2-n}{4pt}}_{\scaleto{n}{2pt}}}{12pt}}}{\frac{2}{n}-1}\bigg)^{2}(1-\epsilon^{4-2n}), & {\rm otherwise}.
\end{cases}
\label{eq:LHS_large_k}
\ee

The Eqs.~(\ref{eq:LHS_large_k}) and (\ref{eq:RHS_large_k}) lead to,
\be
\frac{|{\rm LHS}|}{|{\rm RHS}|}\sim
\begin{cases}
  1+\frac{\ln(\epsilon)}{2\ln({\scaleto{\kappa_{\scaleto{2}{2pt}}}{6pt}})}, & \text{if $n=2$}.\\
  \frac{1}{2}\big(1+\epsilon^{2-n}\big), & {\rm otherwise}.
\end{cases}
\label{eq2:LHS_RHS_small_k}
\ee

Note that for $n \leq 2$, RHS and LHS are positive. Since for $\epsilon < 1$, from Eq.~(\ref{eq2:LHS_RHS_small_k}) we have $|{\rm LHS}| < |{\rm RHS}|$,  it follows that  LHS $<$ RHS. For $n > 2$, RHS and LHS are negative.  In that case, Eq.~(\ref{eq2:LHS_RHS_small_k}) implies $|{\rm LHS}| > |{\rm RHS}|$, and hence again LHS $<$ RHS.

\begin{figure}[hb!]
  \begin{minipage}{.5\textwidth}
    \begin{subfigure}[b]{0.49\textwidth}
      \includegraphics[width = 1.0\textwidth,height=0.14\textheight]{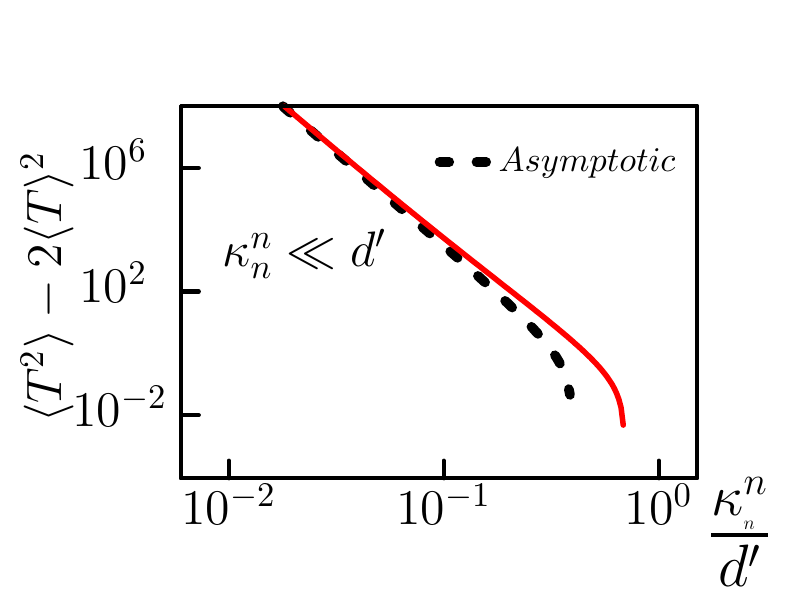}
      \caption{}\label{fig:K_Small_d_9_log_log}
    \end{subfigure}
    \begin{subfigure}[b]{0.49\textwidth}
      \includegraphics[width = 1.0\textwidth,height=0.14\textheight]{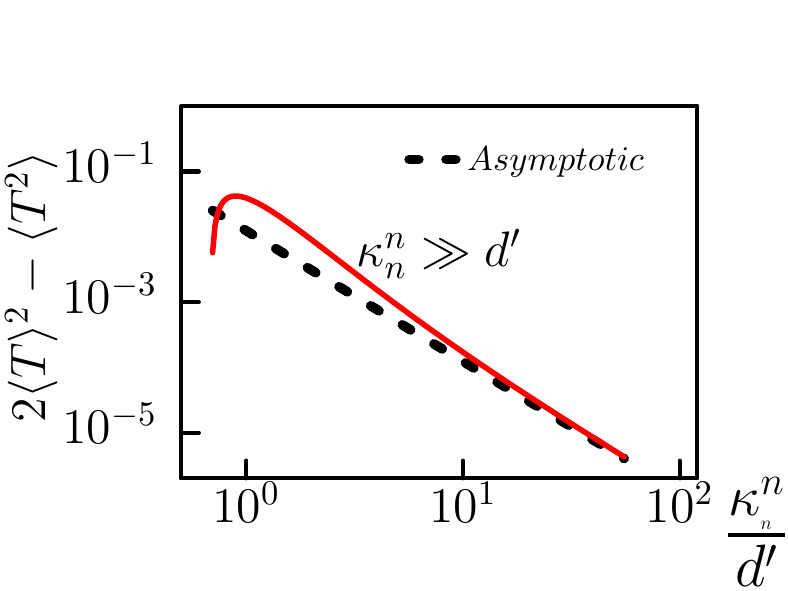}
      \caption{}\label{fig:K_Large_n_3_d_9}
    \end{subfigure}
    \begin{subfigure}[b]{0.99\textwidth}
        \includegraphics[width = 1.0\textwidth,height=0.23\textheight]{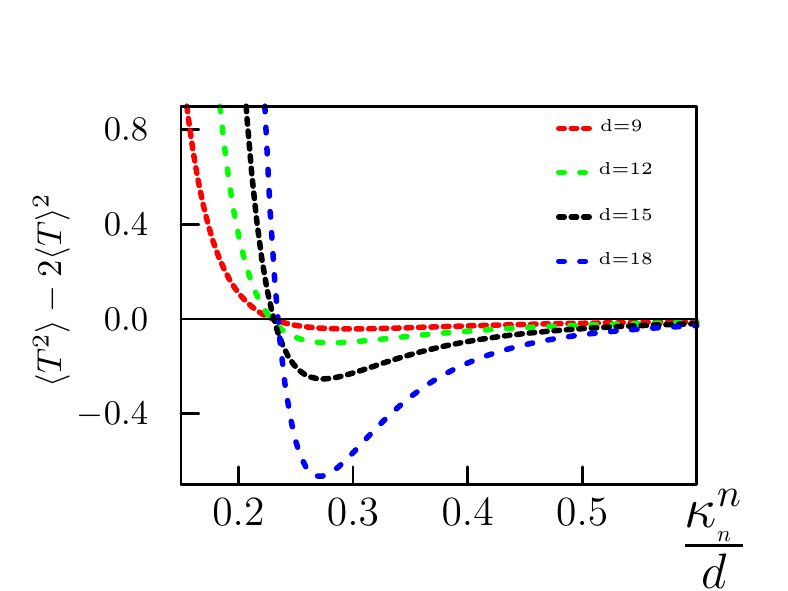}
        \caption{}
        \label{fig:K_Small_Large_d}
    \end{subfigure}
    \caption{In (a) $\langle T^2 \rangle-2\langle T \rangle^2$ is plotted against $\frac{\kappa_n^{n}}{d'} < 1$ and in (b) $2\langle T \rangle^2-\langle T^2 \rangle $ is plotted against $\frac{\kappa_n^{n}}{d'} > 1$ in log-log scale (see solid lines). The corresponding asymptotic values obtained using Eqs.~(\ref{eq:RHS_small_k}), (\ref{eq:LHS_small_k}), (\ref{eq:RHS_large_k}), and (\ref{eq:LHS_large_k}), are shown in dashed lines. Here $n=3$ and dimension $d=9$.  (c) We have shown $\langle T^2 \rangle-2\langle T \rangle^2$ against $\frac{\kappa_n^{n}}{d}$ for $n=3$ for various dimensions $d$. The unique crossing point establishes the dependence of the critical point on $d$. We use $a=0.5$, $R_0=1.0$ and $D=0.5$ for all the sub-figures.}
    \label{fig:Quartic_Pot_Transition1}
  \end{minipage}
  \hfill
  \begin{minipage}{0.4\textwidth}
  \end{minipage}
\end{figure}

Note that $\langle T^2 \rangle - 2 \langle T \rangle^2$ is proportional to (LHS $-$ RHS), and the proportionality constant is known exactly. In Fig.~((\ref{fig:K_Small_d_9_log_log}),(\ref{fig:K_Large_n_3_d_9})), the asymptotic curves of $|\langle T^2 \rangle - 2 \langle T \rangle^2|$ for $\kappa_{n}^n \ll d'$ and $\kappa_n^n \gg d'$ obtained using (LHS $-$ RHS) from Eqs.~(\ref{eq:RHS_small_k}, \ref{eq:LHS_small_k}) and Eqs.~(\ref{eq:RHS_large_k}, \ref{eq:LHS_large_k}) respectively, are compared with their actual values.  The matches are good  slightly away from $\kappa_n^n \sim d'$. The inequalities established in (i) namely LHS $>$ RHS, and (ii) namely LHS $<$ RHS, imply that RHS $=$ LHS in Eq.~(\ref{eq3:d_tansition_n})  when $\kappa_n^n \sim d'$.  In Fig.~(\ref{fig:K_Small_Large_d}) we show that as a consequence, the curves of $\langle T^2 \rangle - 2 \langle T \rangle^2$ for various dimensions $d$ (with fixed $n$) all cross at the same point when plotted against the scaled variable $\kappa_n^n /d$.  Hence we finally have the result that for any $n$ at large $d$, the dimensionless critical point:
\be
\kappa_{c,n}^{\rm asym} \sim d^{\frac{1}{n}}.   
\ee

The asymptotic form ($\kappa_{c,n}^{\rm asym}$) indicates that for the general power law potential, the advantage of resetting remains hard to offset by attractive bias in higher dimensions.


\section{Discussion}
A point of continuous transition in a stochastic system in all dimensions is not always tractable. 
Here we have obtained the exact dimensionless critical strength of a potential 
bias which offsets the advantage rendered by stochastic resetting strategy to first passage of a random walker,
in all dimensions. 
We have done this for different attractive potentials: the logarithmic potential  $k\ln(R/a)$, the uniform box potential, 
and for the general power law potential $V(R)=kR^{n}$ at special $d$ and $n$ (namely when $d/n$ and $2/n$ are integers).
For general $d$ and $n$ even though closed formulas could not be found, we have provided the exact equality (Eq.~(\ref{eq3:d_tansition_n}))
involving some integrals, which may be numerically integrated to obtain the required transition point. Apart from these, for all the above three 
potential functions, the asymptotic dependence of the critical potential on $d$ at large $d$ have been obtained. 
Thus the role of dimensions $d$ is made mathematically explicit in this problem of contemporary interest.  

The qualitative implication of this work is as follows. Resetting confines a random walker to a region of space near the target facilitating quicker capture. Potential bias too pushes the walker to the target making capture happen faster.  The second strategy beyond a threshold strength makes the role of the first strategy of resetting appear as a hindrance. This work shows that multiple directions of escape at higher dimensions provide a better scope for realising the efficacy of the resetting  strategy, even in the face of a bias. To offset this efficacy, the bias strength has to diverge,  
exponentially in fact, as a function of $1/n$ at any given $d$: i.e. $\kappa_{c,n}^{\rm asym} \sim \exp(\frac{1}{n} \ln d)$. Note that $1/n$ is a measure of the steepness of the potential at large $R$ --- smaller the value of $1/n$, the steeper is the potential.  Thus for $n < 1$ (flatter potentials at large $R$) it gets really hard to achieve this transition by tuning bias strength, and thus resetting generally remains a preferred strategy in higher dimensions for such cases.  On the other hand, for $n \gg 1$, it gets much easier to outperform resetting by a strong bias.   

It would be interesting to study the role of dimensions in the cases of discontinuous transitions found earlier in related problems
in which bias and resetting competes. Also the possibility of departure from the hitherto observed mean-field 
paradigm, in different dimensions, remains open for future works.

{\bf Acknowledgement}: We would like to thank A. Nandi, and M. Mitra for useful comments. DD would like to acknowledge SERB India (grant no. MTR/2019/000341) for financial support.

\appendix
\begin{widetext}
  \section{ \label{APP:1} Detailed derivation of the transition point for the Logarithmic Potential.}
  \subsection{\label{Sub:secn=0}\textbf{Obtaining $\nu_c$ using Eq.~(\ref{eq:ORR_n=0})}}
The transcendental equation of ORR $z_*$ from Eq.~(\ref{eq:ORR_n=0}) can be written as
\be
\frac{1-\big(\frac{\epsilon z_*}{2}\big)\epsilon^{\nu}\frac{K_{\nu-1}(\epsilon z_*)}{K_{\nu}(z_*)}}{1-\big(\frac{z_*}{2}\big)\frac{K_{\nu-1}(z_*)}{K_{\nu}(z_*)}}=\epsilon^{\nu}\frac{K_{\nu}(\epsilon z_*)}{K_{\nu}(z_*)}.
\label{eq:App_ORR_n=0}
\ee
For $z_* \to 0$ we use the series expansion $K_{\nu}(x)=\frac{\Gamma \left( \nu \right)}{2}\big(\frac{2}{x}\big)^{\nu}\big[1-\frac{(\frac{x}{2})^2}{(\nu-1)}+\frac{(\frac{x}{2})^4}{2(\nu-1)(\nu-2)}+....\big]$.
Thus, we have
$$\frac{K_{\nu-1}(\epsilon z_*)}{K_{\nu}(z_*)}=\frac{\frac{\Gamma \left( \nu-1 \right)}{2}\big(\frac{2}{\epsilon z_*}\big)^{\nu-1}\big[1-\frac{(\frac{\epsilon z_*}{2})^2}{(\nu-2)}+\frac{(\frac{\epsilon z_*}{2})^4}{2(\nu-2)(\nu-3)}+....\big]}{\frac{\Gamma \left( \nu \right)}{2}\big(\frac{2}{z_*}\big)^{\nu}\big[1-\frac{(\frac{z_*}{2})^2}{(\nu-1)}+\frac{(\frac{z_*}{2})^4}{2(\nu-1)(\nu-2)}+....\big]}$$

$$\implies\big(\frac{\epsilon z_*}{2}\big)\epsilon^{\nu}\frac{K_{\nu-1}(\epsilon z_*)}{K_{\nu}(z_*)}=\frac{\epsilon^2}{\nu-1}\bigg(\frac{z_*}{2}\bigg)^2\bigg(1+\bigg(\frac{1}{\nu-1}-\frac{\epsilon^2}{\nu-2}\bigg)\bigg(\frac{z_*}{2}\bigg)^2+\mathcal{O}\bigg(\frac{z_*}{2}\bigg)^4....\bigg)$$
The right hand part of Eq.(\ref{eq:App_ORR_n=0}) is,
$$\frac{K_{\nu}(\epsilon z_*)}{K_{\nu}(z_*)}=\frac{\frac{\Gamma \left( \nu \right)}{2}\big(\frac{2}{\epsilon z_*}\big)^{\nu}\big[1-\frac{(\frac{\epsilon z_*}{2})^2}{(\nu-1)}+\frac{(\frac{\epsilon z_*}{2})^4}{2(\nu-1)(\nu-2)}+....\big]}{\frac{\Gamma \left( \nu \right)}{2}\big(\frac{2}{z_*}\big)^{\nu}\big[1-\frac{(\frac{z_*}{2})^2}{(\nu-1)}+\frac{(\frac{z_*}{2})^4}{2(\nu-1)(\nu-2)}+....\big]}$$

$$\implies\epsilon^{\nu}\frac{K_{\nu}(\epsilon z_*)}{K_{\nu}(z_*)}=1+\frac{1-\epsilon^2}{\nu-1}\bigg(\frac{z_*}{2}\bigg)^2+\bigg(\frac{1-\epsilon^2}{(\nu-1)^2}-\frac{(1-\epsilon^4)}{2(\nu-1)(\nu-2)}\bigg)\bigg(\frac{z_*}{2}\bigg)^4+\mathcal{O}\bigg(\frac{z_*}{2}\bigg)^6....$$
By putting the above expansions in Eq.~(\ref{eq:App_ORR_n=0}) we get,
\be
\begin{split}
  \frac{1-\frac{\epsilon^2}{\nu-1}\bigg(\frac{z_*}{2}\bigg)^2\bigg(1+\bigg(\frac{1}{\nu-1}-\frac{\epsilon^2}{\nu-2}\bigg)\bigg(\frac{z_*}{2}\bigg)^2+\mathcal{O}\bigg(\frac{z_*}{2}\bigg)^4....\bigg)}{1-\frac{1}{\nu-1}\bigg(\frac{z_*}{2}\bigg)^2\bigg(1+\bigg(\frac{1}{\nu-1}-\frac{1}{\nu-2}\bigg)\bigg(\frac{z_*}{2}\bigg)^2+\mathcal{O}\bigg(\frac{z_*}{2}\bigg)^4....\bigg)}=1+\frac{1-\epsilon^2}{\nu-1}\bigg(\frac{z_*}{2}\bigg)^2\textcolor{white}{--------------}\\
  +\bigg(\frac{1-\epsilon^2}{(\nu-1)^2}-\frac{(1-\epsilon^4)}{2(\nu-1)(\nu-2)}\bigg)\bigg(\frac{z_*}{2}\bigg)^4+\mathcal{O}\bigg(\frac{z_*}{2}\bigg)^6....
\label{eq1:App_ORR_n=0}
\end{split}
\ee
\be
\begin{split}
  1+\frac{1-\epsilon^2}{\nu-1}\bigg(\frac{z_*}{2}\bigg)^2+\bigg(\frac{2(1-\epsilon^2)}{(\nu-1)^2}-\frac{(1-\epsilon^4)}{(\nu-1)(\nu-2)}\bigg)\bigg(\frac{z_*}{2}\bigg)^4+\mathcal{O}\bigg(\frac{z_*}{2}\bigg)^6.... = 1+\frac{1-\epsilon^2}{\nu-1}\bigg(\frac{z_*}{2}\bigg)^2\textcolor{white}{--------}\\
+\bigg(\frac{1-\epsilon^2}{(\nu-1)^2}-\frac{(1-\epsilon^4)}{2(\nu-1)(\nu-2)}\bigg)\bigg(\frac{z_*}{2}\bigg)^4+\mathcal{O}\bigg(\frac{z_*}{2}\bigg)^6....
\label{eq2:App_ORR_n=0}
\end{split}
\ee
After cancelling first and second terms from both sides and further simplification for $z_*\to 0$ leads to the transition point which satisfy
\be
 \frac{2(1-\epsilon^2)}{(\nu_c-1)^2}-\frac{(1-\epsilon^4)}{(\nu_c-1)(\nu-2)} =\frac{1-\epsilon^2}{(\nu_c-1)^2}-\frac{(1-\epsilon^4)}{2(\nu_c-1)(\nu_c-2)}
\label{eq3:App_ORR_n=0}
\ee
Thus
\be
\nu_c=\frac{3-\epsilon^2}{1-\epsilon^2}.
\label{eq:App_ORR_transition_n_0}
\ee
\subsection{\label{Sub1:secn=0}\textbf{Obtaining $\nu_c$ using $\langle T \rangle^2$ and $\langle T^2 \rangle$}}
The Eq.~(\ref{eq:d_mean_time_n=0}) is a second order linear differential equation with boundary condition $\langle T \rangle|_{R=a}=0$. For potential $V(R)=k\ln(R/a)$ it can be transformed into linear first order differential equation for $\xi=\frac{d \langle T \rangle}{d R}$:
\be
\frac{d   \xi}{d R}+\bigg(\frac{1-2\nu}{R}\bigg)\xi=-\frac{1}{D},
\label{eq:d1_mean_time_n=0}
\ee
where $b\equiv \nu (d,k) =1-\big(\frac{d}{2}-\frac{k}{2D}\big)$ depends on both $k$ and $d$.
The solution is $\xi=\frac{R}{2D(\nu-1)}+cR^{(2\nu-1)}$ which leads to $\langle T \rangle$ 
\be
\langle T \rangle=\frac{R^2}{4D(\nu-1)}+c_1\frac{R^{2\nu}}{2\nu}+c_1',
\label{eq:d2_mean_time_n=0}
\ee
where $c_1$ and $c_1'$ are constants. The physically possible case is that, by increasing strength, the mean decrease. However $\nu$ increases, hence for nonzero $c_1$, the $\langle T \rangle$ increases for particular dimension, which contradicting the desired solution. Thus by setting $c_1=0$ and further using absorbing boundary we have,
\be
\langle T \rangle=\frac{R^2-a^2}{4D(\nu-1)}=\frac{R^2(1-\epsilon^2)}{4D(\nu-1)},
\label{eq:d3_mean_time_n=0}
\ee
where $\epsilon=a/R$

Similarly equation of second moment,
\be
\frac{d^2  \langle T^2 \rangle}{d R^2}+\bigg(\frac{1-2\nu}{R}\bigg)\frac{d \langle T^2 \rangle}{d R}=-\frac{R^2-a^2}{2D^2(\nu-1)},
\label{eq:d_2nd_moment_time_n=0}
\ee
Again, if we assume  $\phi=\frac{d \langle T^2 \rangle}{d R}$, we get the solution $\phi=\frac{1}{2D^2(\nu-1)}\big(\frac{a^2R}{2(1-\nu)}-\frac{R^3}{2(2-\nu)}\big)+c_2R^{(2\nu-1)}$. Therefore the second moment is,
\be
\langle T^2 \rangle=\frac{1}{4D^2(\nu-1)}\bigg(\frac{a^2R^2}{2(1-\nu)}-\frac{R^4}{4(2-\nu)}\bigg)+c_2\frac{R^{(2\nu)}}{2\nu}+c'_2.
\label{eq:d1_2nd_moment_time_n=0}
\ee
For the physically possible case $c_2=0$, and applying absorbing boundary condition, we have,
\be
\begin{split}
\langle T^2 \rangle=-\frac{a^2R^2}{8D^2(\nu-1)^2}+\frac{R^4}{16D^2(\nu-1)(\nu-2)}
  +\frac{a^{4}(\nu-3)}{16D^2(\nu-1)^2(\nu-2)},
\label{eq:d2_2nd_moment_time_n=0}
\end{split}
\ee
In terms of dimensionless parameter $\epsilon$,
\be
\langle T^2 \rangle=\frac{R^4(1-\epsilon^2)}{16D^2(\nu-1)^2}\bigg(\frac{(\nu-1)(1+\epsilon^2)-2\epsilon^2(\nu-2)}{\nu-2}\bigg),
\label{eq:d3_2nd_moment_time_n=0}
\ee
The transition happen at when $\langle T^2 \rangle=2\langle T \rangle^2$ (see Eq.~(\ref{eq:transition})) and $\nu\to \nu_c$. Hence we get 
\be
\nu_c=\frac{3-\epsilon^2}{1-\epsilon^2}
\label{eq:transition_Log}
\ee
The dimensionless scaled strength ${\scaleto{\kappa_{\scaleto{c, l}{4pt}}}{10pt}}=\frac{k_c}{D}$ is therefore,
\be
{\scaleto{\kappa_{\scaleto{c, l}{4pt}}}{10pt}}=\frac{4}{1-\epsilon^2}+d
\label{eq:transition_Log1}
\ee
For relatively small absorbing ball i.e ($\epsilon\to 0$); there exist universal $\nu_c=3$ and linear dependence of $\scaleto{\kappa_{\scaleto{c, l}{4pt}}}{10pt}$ on spatial dimension $d$. 

\section{ \label{APP:2}Detailed derivation of the transition point for the Box Potential.}
\subsection{\label{Sub:secn=infty}\textbf{The ORR point for the Box Potential}}
In terms of dimensionless parameter $\epsilon=\frac{a}{R_0}$, ${\scaleto{\kappa_{\scaleto{\infty}{1.3pt}}}{8pt}}=\frac{R_0}{R_m}$ and $z=\alpha_0R_0$ the point $z_*$ is given by: 
\be
\begin{split}
\frac{\big(\frac{z_*\epsilon^{\nu+1}}{2}\big)\bigg(\frac{I_{\nu-1}(\epsilon z_*)K_{\nu-1}(\frac{z_*}{{\scaleto{\kappa_{\scaleto{\infty}{0.8pt}}}{5pt}}})-K_{\nu-1}(\epsilon z_*)I_{\nu-1}(\frac{z_*}{{\scaleto{\kappa_{\scaleto{\infty}{0.8pt}}}{5pt}}})}{I_{\nu}(z_*)K_{\nu-1}(\frac{z_*}{{\scaleto{\kappa_{\scaleto{\infty}{0.8pt}}}{5pt}}})+K_{\nu}(z_*)I_{\nu-1}(\frac{z_*}{{\scaleto{\kappa_{\scaleto{\infty}{0.8pt}}}{5pt}}})}\bigg)-\big(\frac{z_*\epsilon^{\nu}}{2{\scaleto{\kappa_{\scaleto{\infty}{0.8pt}}}{5pt}}}\big)\bigg(\frac{I_{\nu}(\epsilon z_*)K_{\nu}(\frac{z_*}{{\scaleto{\kappa_{\scaleto{\infty}{0.8pt}}}{5pt}}})-K_{\nu}(\epsilon z_*)I_{\nu}(\frac{z_*}{{\scaleto{\kappa_{\scaleto{\infty}{0.8pt}}}{5pt}}})}{I_{\nu}(z_*)K_{\nu-1}(\frac{z_*}{{\scaleto{\kappa_{\scaleto{\infty}{0.8pt}}}{5pt}}})+K_{\nu}(z_*)I_{\nu-1}(\frac{z_*}{{\scaleto{\kappa_{\scaleto{\infty}{0.8pt}}}{5pt}}})}\bigg)+1}{\frac{z_*}{2}\bigg(\frac{I_{\nu-1}( z_*)K_{\nu-1}(\frac{z_*}{{\scaleto{\kappa_{\scaleto{\infty}{0.8pt}}}{5pt}}})-K_{\nu-1}(z_*)I_{\nu-1}(\frac{z_*}{{\scaleto{\kappa_{\scaleto{\infty}{0.8pt}}}{5pt}}})}{I_{\nu}(z_*)K_{\nu-1}(\frac{z_*}{{\scaleto{\kappa_{\scaleto{\infty}{0.8pt}}}{5pt}}})+K_{\nu}(z_*)I_{\nu-1}(\frac{z_*}{{\scaleto{\kappa_{\scaleto{\infty}{0.8pt}}}{5pt}}})}\bigg)-\frac{z_*}{2{\scaleto{\kappa_{\scaleto{\infty}{0.8pt}}}{5pt}}}\bigg(\frac{I_{\nu}(z_*)K_{\nu}(\frac{z_*}{{\scaleto{\kappa_{\scaleto{\infty}{0.8pt}}}{5pt}}})-K_{\nu}(z_*)I_{\nu}(\frac{z_*}{{\scaleto{\kappa_{\scaleto{\infty}{0.8pt}}}{5pt}}})}{I_{\nu}(z_*)K_{\nu-1}(\frac{z_*}{{\scaleto{\kappa_{\scaleto{\infty}{0.8pt}}}{5pt}}})+K_{\nu}(z_*)I_{\nu-1}(\frac{z_*}{{\scaleto{\kappa_{\scaleto{\infty}{0.8pt}}}{5pt}}})}\bigg)+1}\\
=\epsilon^{\nu}\bigg(\frac{I_{\nu}(\epsilon z_*)K_{\nu-1}(\frac{z_*}{{\scaleto{\kappa_{\scaleto{\infty}{0.8pt}}}{5pt}}})+K_{\nu}(\epsilon z_*)I_{\nu-1}(\frac{z_*}{{\scaleto{\kappa_{\scaleto{\infty}{0.8pt}}}{5pt}}})}{I_{\nu}(z_*)K_{\nu-1}(\frac{z_*}{{\scaleto{\kappa_{\scaleto{\infty}{0.8pt}}}{5pt}}})+K_{\nu}(z_*)I_{\nu-1}(\frac{z_*}{{\scaleto{\kappa_{\scaleto{\infty}{0.8pt}}}{5pt}}})}\bigg)
\end{split}
\ee


\subsection{\label{Sub2:secn=infty}\textbf{Study of the transition using $\langle T \rangle$ and  $\langle T^2 \rangle$.}}
For uniform box potential of size $R_m$, the equation moments can be written as
$D\nabla^{2} \langle T \rangle=-1$ and $D\nabla^{2} \langle T^2 \rangle=-2 \langle T \rangle$ with absorbing boundary conditions $\langle T \rangle|_{R=a}=0$, $\langle T^{2} \rangle|_{R=a}=0$,  and reflecting boundary conditions $\frac{d\langle T \rangle}{d R}|_{R=R_m}=0$, $\frac{d\langle T^2 \rangle}{d R}|_{R=R_m}=0$. For isotropic space we have differential equation equivalent to Eq.~(\ref{eq:d_mean_time_n=0}) with solution in similar form as Eq.~(\ref{eq:d2_mean_time_n=0}). For $d\neq 2$ (i.e $\mu\neq 0$),
\be
\langle T \rangle=\frac{R^2}{4D(\mu-1)}+c_3\frac{R^{2\mu}}{2\mu}+c_3',
\label{eq:d1_mean_time_n=infy}
\ee
where index $b\equiv\mu(d) = 1-\frac{d}{2}$ depends on $d$.
After applying boundary conditions we have solution
\be
\langle T \rangle=\frac{R^{2(1-\mu)}_m(a^{2\mu}-R^{2\mu})}{4D\mu(\mu-1)}-\frac{(a^{2}-R^{2})}{4D(\mu-1)},
\label{eq:d2_mean_time_n=Infy}
\ee 
The mean time can be written in terms of dimensionless parameters,
\be
\langle T \rangle=\bigg(\frac{R^2{\scaleto{\kappa^{\scaleto{2(\mu-1)}{5.5pt}}_{\scaleto{\infty}{1.4pt}}}{15pt}}\epsilon^{2\mu}}{4D\mu(\mu-1)}\bigg)\bigg[1-\epsilon^{-2\mu}\big(1+\mu(\epsilon^2-1){\scaleto{\kappa^{\scaleto{2(1-\mu)}{5.5pt}}_{\scaleto{\infty}{1.4pt}}}{15pt}}\big)\bigg],
\label{eq:d2_mean_time_n=Infy}
\ee
Similar to the technique used to find the $\langle T^2 \rangle$ for the log potential, we find the $\langle T^2 \rangle$ for this potential. Here we use reflecting boundary condition $\frac{d \langle T^2 \rangle}{d R}|_{R=R_m}=0$. In terms dimensionless parameter we have
%
\be
\Scale[0.99]
{
\begin{split}
\langle T^2 \rangle=\bigg(\frac{R^4{\scaleto{\kappa^{\scaleto{4(\mu-1)}{5.5pt}}_{\scaleto{\infty}{1.4pt}}}{15pt}}\epsilon^{4\mu}}{8D^2\mu^2(\mu-1)^2}\bigg)\bigg[\mu\epsilon^{-2\mu}{\scaleto{\kappa^{\scaleto{2(1-\mu)}{5.5pt}}_{\scaleto{\infty}{1.4pt}}}{15pt}}(1-\epsilon^2)(1-\mu\epsilon^{2(1-\mu)}{\scaleto{\kappa^{\scaleto{2(1-\mu)}{5.5pt}}_{\scaleto{\infty}{1.4pt}}}{15pt}})+\frac{\mu^2(\mu-1)(1-\epsilon^4)\epsilon^{-4\mu}{\scaleto{\kappa^{\scaleto{4(1-\mu)}{5.5pt}}_{\scaleto{\infty}{1.4pt}}}{15pt}}}{2(\mu-2)}\\
+\frac{\mu(\mu-1)(\epsilon^{-4\mu}-\epsilon^{2(1-\mu)}){\scaleto{\kappa^{\scaleto{2(1-\mu)}{5.5pt}}_{\scaleto{\infty}{1.4pt}}}{15pt}}}{(\mu+1)}+(1-\epsilon^{-2\mu})
\bigg(1+\frac{2(\mu-1)^2{\scaleto{\kappa^{\scaleto{-2\mu}{5.5pt}}_{\scaleto{\infty}{1.4pt}}}{15pt}}\epsilon^{-2\mu}}{\mu-2}-\mu\epsilon^{2(1-\mu)}{\scaleto{\kappa^{\scaleto{2(1-\mu)}{5.5pt}}_{\scaleto{\infty}{1.4pt}}}{15pt}}\bigg)\bigg],
\label{eq:d3_mean_time_n=Infy}
\end{split}}
\ee  
We use the condition of transition Eq.~(\ref{eq:transition}), further equating up to $\epsilon^{-2\mu}$,
\be
\begin{split}
\frac{2(1-\mu)^2}{2-\mu}={\scaleto{\kappa^{\scaleto{2\mu}{5pt}}_{\scaleto{c,\infty}{2.4pt}}}{15pt}}-\bigg(\mu+\mu\bigg(\frac{3+5\mu}{1+\mu}\bigg)\epsilon^2\bigg){\scaleto{\kappa^{\scaleto{2}{3.4pt}}_{\scaleto{c,\infty}{2.4pt}}}{15pt}}+\epsilon^{-2\mu}\bigg(\frac{\mu^2(1-\mu){\scaleto{\kappa^{\scaleto{2(2-\mu)}{5pt}}_{\scaleto{c,\infty}{2.4pt}}}{15pt}}}{4(2-\mu)}+\frac{\mu(1-\mu){\scaleto{\kappa^{\scaleto{2\mu}{4pt}}_{\scaleto{c,\infty}{2pt}}}{15pt}}}{1+\mu}\\
-\frac{2(1-\mu)^2}{2-\mu}+({\scaleto{\kappa^{\scaleto{\mu}{3pt}}_{\scaleto{c, \infty}{2.4pt}}}{15pt}}-\mu{\scaleto{\kappa^{\scaleto{-2\mu}{5pt}}_{\scaleto{c,\infty}{2.4pt}}}{15pt}})^2\bigg)
\label{eq:transition_box_appen}
\end{split}
\ee
Since for $\epsilon<1$ we can neglect $\epsilon^{-2\mu}$ in higher dimension. Moreover if we neglect $\epsilon^2$ term we get Eq.~(\ref{eq:transition_box}). 
 
In terms of dimension $d$ the  Eq.~(\ref{eq:transition_box}) can be written as: 
\be
{\scaleto{\kappa^{\scaleto{d-2}{4pt}}_{\scaleto{c,\infty}{2.4pt}}}{18pt}}=\frac{d+2}{d^2}\bigg(1+\bigg(\frac{d-2}{2}\bigg){\scaleto{\kappa^{\scaleto{d}{4pt}}_{\scaleto{c,\infty}{2.4pt}}}{18pt}}\bigg)
\label{eq1:transition_box}
\ee
It can be numerically shown that the value of ${\scaleto{\kappa_{\scaleto{c,\infty}{2.4pt}}}{10pt}}$ is very close to $1$ for high $d$. Thus we define $\delta = 1-{\scaleto{\kappa_{\scaleto{c,\infty}{2.4pt}}}{10pt}}$ as very small quantity. We get 
\be
e^{-\delta(d-2)}=\frac{d+2}{d^2}\bigg(1+\bigg(\frac{d-2}{2}\bigg)e^{-\delta d}\bigg)
\label{eq2:transition_box}
\ee
Taking small series expansion and further simplifying we have
\be
e^{\delta d} =\frac{1+\frac{4\delta(\delta+1)d^2}{d^2+4}}{\frac{4+2d}{4+d^2}}. 
\label{eq4:transition_box}
\ee
Thus for large $d$, we have $\delta d=\ln(1+4\delta(\delta+1))+\ln(d/2)$. For small $\delta$ 
we have 
\be
\delta (d-4)=\ln(d/2)-4\delta^2
\label{eq5:transition_box}
\ee
hence the final result (Eq.~(\ref{eq:Box_Pot_Large_d})) with an additional higher order term is   
\be
\delta =\frac{\ln(d/2)}{d}-4\frac{(ln(d/2))^2}{d^3}
\label{eq6:transition_box}
\ee

\section{ \label{APP:3} Detailed derivations related to power-law potentials $V(R)=kR^n$.}
\subsection{\label{App3Sub1:secn=n_transition}\textbf{Derivation of the transition condition --- Eq.~(\ref{eq3:d_tansition_n})}}
For potential $V(R)=kR^n$ with $k>0$ and  integer $n>0$, the equation of mean time $\langle T \rangle$ from Eq.~(\ref{eq:1st_Moment_without_r}) can be written as
\be
\frac{d^2  \langle T \rangle}{d R^2}+\bigg(\frac{d-1}{R}-n\bigg(\frac{k}{D}\bigg)R^{n-1}\bigg)\frac{d \langle T \rangle}{d R}=-\frac{1}{D},
\label{App_eq:d_mean_time_n}
\ee
Assuming $\xi=\frac{d \langle T \rangle}{d R}$, we get linear first order differential equation with solution;
\be
\frac{d \xi}{d R}+\bigg(\frac{d-1}{R}-n\bigg(\frac{k}{D}\bigg)R^{n-1}\bigg) \xi=-\frac{1}{D},
\label{App_eq1:d_mean_time_n}
\ee
\be
\xi=\frac{R^{1-d}e^{\frac{k}{D}R^n}}{D}\int^{\infty}_{R} R'^{d-1} e^{-\frac{k}{D}R'^n} dR'.
\label{App_eq2:d_mean_time_n}
\ee
Using the scaled variable $x=\frac{kR^n}{D}$,
\be
\xi=\frac{R^{1-d}e^{\frac{k}{D}R^n}}{nD}\bigg(\frac{k}{D}\bigg)^{-d'}\int^{\infty}_{\frac{kR^n}{D}} x^{d'-1} e^{-x} dx,
\label{App_eq3:d_mean_time_n}
\ee
\be
\xi=\frac{R}{nD}\bigg(\frac{kR^n}{D}\bigg)^{-d'}e^{\frac{kR^n}{D}}\Gamma\bigg(d',\frac{kR^n}{D}\bigg),
\label{App_eq4:d_mean_time_n}
\ee
where $d'=\frac{d}{n}$ can be any positive rational number. Thus the moment can be found by integrating from $a$ to $R$
\be
\begin{split}
  \langle T \rangle =\int^{R}_{a}\xi(R')dR'=\frac{1}{n^2D}\bigg(\frac{D}{k}\bigg)^{\frac{2}{n}}\int^{\frac{kR^n}{D}}_{\frac{ka^n}{D}} x^{\frac{2}{n}-d'-1} e^{x}\Gamma(d',x) dx\\
  \equiv\frac{1}{n^2D}\bigg(\frac{D}{k}\bigg)^{\frac{2}{n}}\psi_{1}(d',n,\kappa_n^n)-\phi_{1}(d',n,\epsilon^n \kappa_n^n),
  \label{App_eq5:d_mean_time_n}
  \end{split}
\ee
where by definition $\kappa_{n}^{n} = \frac{k{R}^n}{D}$, and $\frac{k{a}^n}{D} = \epsilon^n \kappa_n^n$ and $\psi_{1}(d',n,x)$ is the indefinite integral $\int x^{\frac{2}{n}-d'-1} e^{x} \Gamma(d', x)dx$. Thus $\psi_1(d',n,\kappa_n^n)$ is to be treated as function of the variable $R$ through $\kappa_n^n$, while $\phi_{1}$ is a constant for given values of $k$, $n$, $d$ and $a$.

The equation of the second moment $\langle T^2 \rangle$ can be written for the potential $V(R)$ as 
\be
\frac{d^2  \langle T^2 \rangle}{d R^2}+\bigg(\frac{d-1}{R}-n\bigg(\frac{k}{D}\bigg)R^{n-1}\bigg)\frac{d \langle T^2 \rangle}{d R}=-\frac{2\langle T \rangle}{D},
\label{App_eq1:d_seconed_mean_time_n}
\ee
Noting that the Eqs~(\ref{App_eq:d_mean_time_n}) and (\ref{App_eq1:d_seconed_mean_time_n}) are of the same form except for the right hand side, we have: 
\be
\frac{d \langle T^2 \rangle}{d R}= \frac{2}{D}R^{1-d}e^{\frac{k}{D}R^n}\int^{\infty}_{R} \langle T \rangle R'^{d-1} e^{-\frac{kR'^n}{D}} dR',
\label{App_eq2:d_second_mean_time_n}
\ee
which on substituting $\langle T \rangle$ (from Eq.~(\ref{App_eq5:d_mean_time_n})) yields,
\be
\frac{d \langle T^2 \rangle}{d R}= \frac{2}{n^2D^2}\bigg(\frac{D}{k}\bigg)^{\frac{2}{n}}R^{1-d}e^{\frac{k}{D}R^n}\int^{\infty}_{R} \psi_{1}\big(d',n, \frac{kR'^{n}}{D} \big) R'^{d-1} e^{-\frac{kR'^n}{D}} dR'.
\label{App_eq3:d_second_mean_time_n}
\ee
Thus by putting $x=\frac{kR^n}{D}$, we have
\be
\frac{d \langle T^2 \rangle}{d R}= \frac{2}{n^3D^2}\bigg(\frac{D}{k}\bigg)^{d'+\frac{2}{n}}R^{1-d}e^{\frac{k}{D}R^n}\int^{\infty}_{\frac{kR^n}{D}} \psi_{1}\big(d',n, x \big) x^{d'-1} e^{-x} dx,
\label{App_eq4:d_second_mean_time_n}
\ee
\be
 \frac{d \langle T^2 \rangle}{d R}= \frac{2}{n^3D^2}\bigg(\frac{D}{k}\bigg)^{d'+\frac{2}{n}} R^{1-d}e^{\frac{kR^n}{D}}\psi_{2}(d',n,\kappa_n^n)-2\phi_{1}(d',n,\epsilon^n \kappa_n^n)\xi(R),
\label{App_eq5:d_seconed_mean_time_n}
\ee
where $\psi_{2}(d',n,x)=\int^{\infty}_{x}x^{'(d'-1)}e^{-x'}\psi_{1}(d',n,x')dx'$.
The $\langle T^2 \rangle$ can be found by integrating from $a$ to $R$,
\be
 \langle T^2 \rangle= \frac{2}{n^3D^2}\bigg(\frac{D}{k}\bigg)^{d'+\frac{2}{n}}\int^{R}_{a} R'^{1-d}e^{\frac{kR'^n}{D}}\psi_{2}(d',n,\frac{kR'^{n}}{D})dR'-2\phi_{1}(d',n,\epsilon^n \kappa_n^n)\xi(R),
\label{App_eq6:d_seconed_mean_time_n}
\ee
Again by putting $x=\frac{kR^n}{D}$, we obtained
\be
 \langle T^2 \rangle =  \frac{2}{n^4D^2}\bigg(\frac{D}{k}\bigg)^{\frac{4}{n}}\int_{\epsilon^n \kappa_n^n}^{\kappa_n^n}x^{\frac{2}{n}-d'-1}\psi_{2}(d',n,x)e^{x}dx-2\phi_{1}(d',n,\epsilon^n \kappa_n^n)\langle T \rangle. 
\label{App_eq7:d_seconed_mean_time_n}
\ee
Using Eqs~(\ref{App_eq5:d_mean_time_n}) and (\ref{App_eq7:d_seconed_mean_time_n}) in Eq.~(\ref{eq:transition}) for the transition point, we finally obtain the equation 
satisfied by the dimensionless critical potential strength $\kappa_{c,n}$: 
\be
 \int_{\epsilon^n {\scaleto{\kappa^{n}_{\scaleto{c, n}{2pt}}}{10pt}}}^{{\scaleto{\kappa^{n}_{\scaleto{c, n}{2pt}}}{10pt}}}x^{\frac{2}{n}-d'-1}\psi_{2}(d',n,x)e^{x}dx=\psi_{1}(d',n,{\scaleto{\kappa^{n}_{\scaleto{c, n}{2pt}}}{10pt}}) \int_{\epsilon^n {\scaleto{\kappa^{n}_{\scaleto{c, n}{2pt}}}{10pt}}}^{{\scaleto{\kappa^{n}_{\scaleto{c, n}{2pt}}}{10pt}}}x^{\frac{2}{n}-d'-1}\Gamma(d',x)e^{x}dx
\label{App_eq3:d_tansition_n}
\ee

\subsection{\label{App3Sub2:secn=n_int_trans}\textbf{Study of the transition for integer value of $d'=\frac{d}{n}$ and $\frac{2}{n}$}}
The gamma function has series expansion $\Gamma (l,x)= \Gamma (l) e^{-x}\sum_{i=0}^{l-1}\frac{x^i}{i!}$ for positive integer $l$.
The Eq.~(\ref{App_eq3:d_mean_time_n}) is exactly integrable only for integer value of $d'=\frac{d}{n}$, which is
\be
\xi=\frac{\Gamma(d')}{nD}\bigg(\frac{k}{D}\bigg)^{-d'}\sum_{i=0}^{d'-1}\bigg(\frac{k}{D}\bigg)^{i}\frac{R^{1-nd'+ni}}{i!},
\label{App_eq1:d_mean_time_integer_dp}
\ee
Thus the moment can be found by integrating from $a$ to $R$

\be
\begin{split}
  \langle T \rangle =\frac{\Gamma(d')\big(\frac{k}{D}\big)^{-d'}}{nD}\bigg(\sum_{\substack{i=0 \\ i\neq d'-\frac{2}{n}}}^{d'-1}\frac{\big(\frac{k}{D}\big)^{i}(R^{2-n(d'-i)}-a^{2-n(d'-i)})}{i!(2-n(d'-i))}+\frac{\big(\frac{k}{D}\big)^{i}\ln\big(\frac{R}{a}\big)}{i!}\bigg|_{i=d'-\frac{2}{n}}\bigg)\\
  \equiv \frac{1}{n^2D}\bigg(\frac{D}{k}\bigg)^{\frac{2}{n}}\psi_{1}(d',n,\frac{kR^n}{D})-\phi_{1}(d',n,\frac{ka^n}{D}),
\label{App_eq2:d_mean_time_integer_dp}
\end{split}
\ee
we get the solution of first derivative 
\be
\frac{d \langle T^2 \rangle}{d R}=\frac{2}{D}R^{1-d}e^{\frac{k}{D}R^n}\int^{\infty}_{R}  \langle T \rangle R'^{d-1} e^{-\frac{k}{D}R'^n} dR',
\label{App_eq1:d_seconed_mean_time_integer_dp}
\ee
or
\be
\Scale[1.0]
{\begin{split}
    \frac{d \langle T^2 \rangle}{d R}=\frac{2\Gamma(d')\big(\frac{k}{D}\big)^{-d'}}{nD^2}R^{1-d}e^{\frac{k}{D}R^n}\bigg(\sum_{\substack{i=0 \\ i\neq d'-\frac{2}{n}}}^{d'-1}\frac{\big(\frac{k}{D}\big)^{i}}{i!(2-n(d'-i))}\int^{\infty}_{R} R'^{(ni+1)} e^{-\frac{k}{D}R'^n} dR'\\
    +\frac{\big(\frac{k}{D}\big)^{i}}{i!}\bigg|_{i=d'-\frac{2}{n}}\int^{\infty}_{R} \ln(R') R'^{d-1} e^{-\frac{k}{D}R'^n} dR'\bigg)-2\phi_{1}(d',n,\frac{ka^n}{D})\xi(R),
    \label{App_eq2:d_seconed_mean_time_integer_dp}
 \end{split}}
\ee
or
\be
\Scale[1.0]
{\begin{split}
    \frac{d \langle T^2 \rangle}{d R}=\frac{2\Gamma(d')\big(\frac{k}{D}\big)^{-d'}}{n^2D^2}R^{1-d}e^{\frac{k}{D}R^n}\bigg(\sum_{\substack{i=0 \\ i\neq d'-\frac{2}{n}}}^{d'-1}\frac{\big(\frac{k}{D}\big)^{-\frac{2}{n}}}{i!(2-n(d'-i))}\Gamma\bigg(\frac{2}{n}+i,\frac{k}{D}R^n\bigg)+\frac{\big(\frac{k}{D}\big)^{i-d'}}{i!}\bigg|_{i=d'-\frac{2}{n}}\bigg(\ln(R)\Gamma\big(d',\frac{k}{D}R^n\big)\\
    +\frac{\Gamma(d')}{n}\sum_{j=0}^{d'-1}\frac{\Gamma(j,\frac{k}{D}R^n)}{j!}\bigg)\bigg)-2\phi_{1}(d',n,\frac{ka^n}{D})\xi(R).
    \label{App_eq3:d_seconed_mean_time_integer_dp}
 \end{split}}
\ee
The Eq.(\ref{App_eq3:d_seconed_mean_time_integer_dp}) is in closed form for an integer $d'$. The solution of $\langle T^2 \rangle$ can be found by integrating Eq.(\ref{App_eq3:d_seconed_mean_time_integer_dp}) from $a$ to $R$. However, numerical integration can only be done for integer $d'$ and any positive integer $n$. 
We note that the analytical integration is  possible for integer `$i+\frac{2}{n}$' hence only for integer values of both $d'$ and $\frac{2}{n}$. 
We proceed further to  find closed form analytical solutions by considering $\frac{2}{n}$ an integer.
\be
\Scale[.90]
{\begin{split}
    \langle T^2 \rangle =\frac{2\Gamma(d')\big(\frac{k}{D}\big)^{-d'}}{n^2D^2}\bigg(\sum_{\substack{i=0 \\ i\neq d'-\frac{2}{n}}}^{d'-1}\frac{\Gamma\big(\frac{2}{n}+i\big)\big(\frac{k}{D}\big)^{-\frac{2}{n}}}{i!(2-n(d'-i))}\bigg[\sum_{\substack{l=0 \\ l\neq d'-\frac{2}{n}}}^{\frac{2}{n}+i-1}\frac{\big(\frac{k}{D}\big)^{l}(R^{2-n(d'-l)}-a^{2-n(d'-l)})}{l!(2-n(d'-l))}+\frac{\big(\frac{k}{D}\big)^{l}\ln(\frac{R}{a})}{l!}\bigg|_{l=d'-\frac{2}{n}}\bigg]\textcolor{white}{--------}\\+\frac{\Gamma(d')\big(\frac{k}{D}\big)^{i-d'}}{i!}\bigg|_{i=d'-\frac{2}{n}}\bigg[\sum_{\substack{m=0 \\ m\neq d'-\frac{2}{n}}}^{d'-1}\frac{\big(\frac{k}{D}\big)^{m}}{m!(2-n(d'-m))}\bigg(R^{2-n(d'-m)}\ln(R)-a^{2-n(d'-m)}\ln(a)-\bigg(\frac{R^{2-n(d'-m)}-a^{2-n(d'-m)}}{2-n(d'-m)}\bigg)\bigg)\\+\frac{\big(\frac{k}{D}\big)^{m}}{2m!}\bigg|_{m=d'-\frac{2}{n}}(\ln^2(R)-\ln^2(a))+\sum_{j=1}^{d'-1}\frac{1}{nj}\sum_{\substack{u=0 \\ u\neq d'-\frac{2}{n}}}^{j-1}\frac{\big(\frac{k}{D}\big)^{u}(R^{2-n(d'-u)}-a^{2-n(d'-u)})}{u!(2-n(d'-u))}+\frac{\big(\frac{k}{D}\big)^{u}\ln(\frac{R}{a})}{u!}\bigg|_{u=d'-\frac{2}{n}}\\
    -\frac{1}{n}\int_{a}^{R}Ei\big(-\frac{k}{D}R'^n\big)R'^{1-d}e^{\frac{k}{D}R'^n}dR'\bigg]\bigg)-2\phi_{1}(d',n,\frac{ka^n}{D})\langle T \rangle,
    \label{App_eq4:d_seconed_mean_time_integer_dp}
 \end{split}}
\ee
where $Ei(x)$ is exponential integral. Thus we have
\be
\Scale[.90]
{\begin{split}
    \langle T^2 \rangle =\frac{2\Gamma(d')\big(\frac{k}{D}\big)^{-d'}}{n^2D^2}\bigg(\sum_{\substack{i=0 \\ i\neq d'-\frac{2}{n}}}^{d'-1}\frac{\Gamma\big(\frac{2}{n}+i\big)\big(\frac{k}{D}\big)^{-\frac{2}{n}}}{i!(2-n(d'-i))}\bigg[\sum_{\substack{l=0 \\ l\neq d'-\frac{2}{n}}}^{\frac{2}{n}+i-1}\frac{\big(\frac{k}{D}\big)^{l}(R^{2-n(d'-l)}-a^{2-n(d'-l)})}{l!(2-n(d'-l))}+\frac{\big(\frac{k}{D}\big)^{l}\ln(\frac{R}{a})}{l!}\bigg|_{l=d'-\frac{2}{n}}\bigg]\textcolor{white}{--------}\\+\frac{\Gamma(d')\big(\frac{k}{D}\big)^{i-d'}}{i!}\bigg|_{i=d'-\frac{2}{n}}\bigg[\sum_{\substack{m=0 \\ m\neq d'-\frac{2}{n}}}^{d'-1}\frac{\big(\frac{k}{D}\big)^{m}}{m!(2-n(d'-m))}\bigg(R^{2-n(d'-m)}\ln(R)-a^{2-n(d'-m)}\ln(a)-\bigg(\frac{R^{2-n(d'-m)}-a^{2-n(d'-m)}}{2-n(d'-m)}\bigg)\bigg)\\+\frac{\big(\frac{k}{D}\big)^{m}}{2m!}\bigg|_{m=d'-\frac{2}{n}}(\ln^2(R)-\ln^2(a))+\sum_{j=1}^{d'-1}\frac{1}{nj}\sum_{\substack{u=0 \\ u\neq d'-\frac{2}{n}}}^{j-1}\frac{\big(\frac{k}{D}\big)^{u}(R^{2-n(d'-u)}-a^{2-n(d'-u)})}{u!(2-n(d'-u))}+\frac{\big(\frac{k}{D}\big)^{u}\ln(\frac{R}{a})}{u!}\bigg|_{u=d'-\frac{2}{n}}\\
      -\frac{1}{n^2}\bigg(R^{2-d}G_{2,3}^{3,1}\bigg(\frac{kR^n}{D}\bigg| \begin{split} {\scaleto{0, d'-\frac{2}{n}+1}{13.4pt}} \\{\scaleto{0, 0, d'-\frac{2}{n}}{13.4pt}}\end{split}\bigg)-a^{2-d}G_{2,3}^{3,1}\bigg(\frac{ka^n}{D}\bigg| \begin{split} {\scaleto{0, d'-\frac{2}{n}+1}{13.4pt}} \\{\scaleto{0, 0, d'-\frac{2}{n}}{13.4pt}}\end{split}\bigg) \bigg)\bigg)\bigg]\bigg)
    -2\phi_{1}(d',n,\frac{ka^n}{D})\langle T \rangle,
    \label{App_eq5:d_seconed_mean_time_integer_dp}
 \end{split}}
\ee
By putting  Eq.~(\ref{App_eq2:d_mean_time_integer_dp}) and Eq.~(\ref{App_eq5:d_seconed_mean_time_integer_dp})  in Eq.~(\ref{eq:transition}) we can find the critical strength analytically 
exactly.

\subsection{\label{App3Sub3:secn=n_transition}\textbf{Detailed study of the transition for any integer $n$ and large $d$}}
In this section using Eq.~(\ref{eq3:d_tansition_n}) we find the transition point in large $d$. We establish that in the limit ${\scaleto{\kappa^{n}_{\scaleto{n}{2pt}}}{12pt}}\ll d'$,  LHS $>$ RHS (where they refer to Eq.~(\ref{eq3:d_tansition_n})). Similarly for ${\scaleto{\kappa^{n}_{\scaleto{n}{2pt}}}{12pt}}\gg d'$ the opposite is true. We find below $\psi_{1}(d',n,x)$ and $\psi_{2}(d',n,x)$ and consequently LHS, and RHS.


 The definition of $\psi_{1}(d',n,x)=\int x^{\frac{2}{n}-d'-1} e^{x} \Gamma(d', x)$ and $\psi_{2}(d',n,x)=\int^{\infty}_{x}x^{'(d'-1)}e^{-x'}\psi_{1}(d',n,x')$ are valid for any $n$ and $d$.  The first thing we note is that 
\be
\Gamma(d',x)\sim
\begin{cases}
    \Gamma(d'), & \text{if $x\ll d'$}.\\
    x^{d'-1}e^{-x}, & \text{if $x\gg d'$}.
\end{cases}
\ee
(i) For ${\scaleto{\kappa^{n}_{\scaleto{n}{2pt}}}{12pt}}\ll d'$ limit:
$$\psi_{1}(d',n,x)\sim \Gamma(d')\int x^{\frac{2}{n}-d'-1}dx \sim - \Gamma(d')\frac{x^{\frac{2}{n}-d'}}{d' - \frac{2}{n}}$$
Similarly using small $x$, 
$$\psi_{2}(d',n,x)\sim -\frac{\Gamma(d')}{d'-\frac{2}{n}}\int_{x}^{\infty}e^{-x'}x'^{(\frac{2}{n}-1)}dx' \sim -\frac{\Gamma(d')\Gamma\big(\frac{2}{n}\big)}{d'-\frac{2}{n}}$$
thus
\be
{\rm RHS}\sim -\bigg(\frac{\Gamma(d'){\scaleto{\kappa^{\scaleto{2-d}{4pt}}_{\scaleto{n}{2pt}}}{12pt}}}{d'-\frac{2}{n}}\bigg)^{2}(\epsilon^{2-d}-1)
\label{App_eq1:RHS_small_k}
\ee
and
\be
{\rm LHS}\sim  -\frac{\Gamma(d')\Gamma\big(\frac{2}{n}\big)}{\big(d'-\frac{2}{n}\big)}\int^{{\scaleto{\kappa^{n}_{\scaleto{n}{2pt}}}{8pt}}}_{\epsilon^n{\scaleto{\kappa^{n}_{\scaleto{n}{2pt}}}{8pt}}} x^{\frac{2}{n}-d'-1} dx= -\frac{\Gamma(d')\Gamma\big(\frac{2}{n}\big)}{\big(d'-\frac{2}{n}\big)^2}{\scaleto{\kappa^{\scaleto{2-d}{4pt}}_{\scaleto{n}{2pt}}}{12pt}}(\epsilon^{2-d}-1)
\label{App_eq1:LHS_small_k}
\ee
So both LHS and RHS are negative for this limit. The ratio is,
\be
\frac{\rm LHS}{\rm RHS}\sim\frac{\Gamma\big(\frac{2}{n}\big)}{\Gamma(d'){\scaleto{\kappa^{\scaleto{2-d}{4pt}}_{\scaleto{n}{2pt}}}{12pt}}}\approx \frac{1}{\rm A}
\label{App_eq1:LHS_RHS_small_k}
\ee
where $\Gamma(2/n) A=\Gamma(d'){\scaleto{\kappa^{\scaleto{2-d}{4pt}}_{\scaleto{n}{2pt}}}{12pt}}\approx {\scaleto{\kappa^{2}_{\scaleto{n}{2pt}}}{12pt}}\exp[d' \ln \big(\frac{d'}{e\kappa_n^n}\big)] \gg 1$ at large $d$ and in limit ${\scaleto{\kappa^{n}_{\scaleto{n}{2pt}}}{8pt}}\ll d'$. Therefore, RHS is more negative than LHS, hence LHS $>$ RHS for ${\scaleto{\kappa^{n}_{\scaleto{n}{2pt}}}{8pt}}\ll d'$.

(ii) For ${\scaleto{\kappa^{n}_{\scaleto{n}{2pt}}}{8pt}}\gg d'$, although $\epsilon<1$, we need to assume $\epsilon^{n}{\scaleto{\kappa^{n}_{\scaleto{n}{2pt}}}{8pt}}\gg d'$. Using form of $\Gamma(d',x)$ for $x\gg d'$ we find values
$$\psi_{1}(d',n,x)\sim\int x^{\frac{2}{n}-2}dx \sim 
\begin{cases}
   \ln(x) , & \text{if $n=2$}.\\
     \frac{x^{\frac{2}{n}-1}}{\frac{2}{n}-1}, & \text{$otherwise$},
\end{cases}$$

and 

$$\psi_{2}(d',n,x)\sim
\begin{cases}
    \int_{x}^{\infty}x'^{d'-1}\ln(x')e^{-x'}dx'\approx x^{d'-1}e^{-x}\ln(x)\big(1+\frac{1+(d'-1)\ln(x)}{x\ln(x)}+\mathcal{O}(\frac{1}{x^2})+....\big), & \text{if $n=2$}.\\
     \int_{x}^{\infty}\frac{x'^{(d'+\frac{2}{n}-2)}e^{-x'}}{\frac{2}{n}-1}dx'=\frac{\Gamma(d'+\frac{2}{n}-1,x)}{\frac{2}{n}-1}, & \text{$otherwise$}.
\end{cases}$$
Thus for large $x \gg d'$, we have:
$$\psi_{2}(d',n,x)\sim
\begin{cases}
   x^{d'-1}e^{-x}\ln(x) , & \text{if $n=2$}.\\
     \frac{x^{d'+\frac{2}{n}-2}e^{-x}}{\frac{2}{n}-1}, & \text{$otherwise$}.
\end{cases}$$
This implies
\be
{\rm RHS}\sim
\begin{cases}
  \ln({\scaleto{\kappa^{2}_{\scaleto{2}{3.5pt}}}{11pt}})(\ln({\scaleto{\kappa^{2}_{\scaleto{2}{3.5pt}}}{11pt}})-\ln(\epsilon^2{\scaleto{\kappa^{2}_{\scaleto{2}{3.5pt}}}{11pt}}))=4\ln({\scaleto{\kappa_{\scaleto{2}{3.pt}}}{7pt}})\ln(1/\epsilon) , & \text{if $n=2$}.\\
    \bigg(\frac{{\scaleto{\kappa^{\scaleto{2-n}{4pt}}_{\scaleto{n}{2.5pt}}}{11pt}}}{\frac{2}{n}-1}\bigg)\bigg(\frac{{\scaleto{\kappa^{\scaleto{2-n}{4pt}}_{\scaleto{n}{2pt}}}{12pt}}-(\epsilon{\scaleto{\kappa_{\scaleto{n}{2pt}}}{6pt}})^{2-n}}{\frac{2}{n}-1}\bigg)=\bigg(\frac{{\scaleto{\kappa^{\scaleto{2-n}{4pt}}_{\scaleto{n}{2pt}}}{12pt}}}{\frac{2}{n}-1}\bigg)^{2}(1-\epsilon^{2-n}), & \text{$otherwise$}
\end{cases}
\label{App_eq1:RHS_large_k}
\ee

and 

$$
{\rm LHS}\sim
\begin{cases}
   \int^{{\scaleto{\kappa^{n}_{\scaleto{n}{2pt}}}{8pt}}}_{\epsilon^n{\scaleto{\kappa^{n}_{\scaleto{n}{2pt}}}{8pt}}}\frac{\ln(x)}{x}dx=\frac{(\ln({\scaleto{\kappa^{2}_{\scaleto{2}{2pt}}}{8pt}}))^{2}-(\ln(\epsilon^2{\scaleto{\kappa^{2}_{\scaleto{n}{2pt}}}{8pt}}))^2}{2}, & \text{if $n=2$}.\\
     \int^{{\scaleto{\kappa^{n}_{\scaleto{n}{2pt}}}{8pt}}}_{\epsilon^n{\scaleto{\kappa^{n}_{\scaleto{n}{2pt}}}{8pt}}}\frac{x^{\frac{4}{n}-3}}{\frac{2}{n}-1}dx=\bigg(\frac{1}{\frac{2}{n}-1}\bigg)\bigg(\frac{{\scaleto{\kappa^{\scaleto{4-n}{4pt}}_{\scaleto{n}{2pt}}}{12pt}}-(\epsilon {\scaleto{\kappa_{\scaleto{n}{2pt}}}{6pt}})^{4-n}}{\frac{4}{n}-2}\bigg), & \text{$otherwise$}
\end{cases}
$$

which simplifies to 
\be
{\rm LHS}\sim 
\begin{cases}
   4\ln({\scaleto{\kappa_{\scaleto{2}{2pt}}}{8pt}})\ln(1/\epsilon)\bigg(1+\frac{\ln(\epsilon)}{2\ln({\scaleto{\kappa_{\scaleto{2}{2pt}}}{6pt}})}\bigg), & \text{if $n=2$}.\\
    \frac{1}{2}\bigg(\frac{{\scaleto{\kappa^{\scaleto{2-n}{4pt}}_{\scaleto{n}{2pt}}}{12pt}}}{\frac{2}{n}-1}\bigg)^{2}(1-\epsilon^{4-2n}), & \text{$otherwise$}.
\end{cases}
\label{App_eq:LHS_large_k}
\ee
Taking the ratio of moduli of  Eqs~(\ref{App_eq:LHS_large_k}) and (\ref{App_eq1:RHS_large_k}), we get 
 \be
\frac{\rm |LHS|}{\rm |RHS|}\sim
\begin{cases}
 1+\frac{\ln(\epsilon)}{2\ln({\scaleto{\kappa_{\scaleto{2}{2pt}}}{6pt}})}, & \text{if $n=2$}.\\
    \frac{1}{2}\big(1+\epsilon^{2-n}\big), & \text{$otherwise$}.
\end{cases}
\label{App_eq1:LHS_RHS_small_k}
\ee
Note that for $n\leq 2$ both LHS and RHS have positive values. Since $\epsilon<1$, implies $\rm |LHS|$ $<$ $\rm |RHS|$ and therefore LHS$<$RHS. For $n>2$ both LHS and RHS are negative but $\rm |LHS|$ $>$ $\rm |RHS|$, and therefore again LHS $<$ RHS.

\end{widetext}
\bibliography{article_2_v3}
\end{document}